\documentclass[english,aps, prc, manuscript,superscriptaddress,reprint,showpacs,amsmafixth,amssymb,float]{revtex4-1}
\usepackage{mathptmx}

\usepackage[T1]{fontenc}
\usepackage[latin9]{inputenc}
\usepackage{geometry}
\usepackage{array}
\usepackage{bm}
\usepackage{graphicx}
\usepackage{esint}
\usepackage[dvipdfm,bookmarks=true,colorlinks,%
            citecolor=blue,linkcolor=blue,hypertex, %
            breaklinks=true]{hyperref}
\makeatletter

\providecommand{\tabularnewline}{\\}
\newcommand{\lyxdot}{.}

 \@ifundefined{textcolor}{}
 {%
   \definecolor{BLACK}{gray}{0}
   \definecolor{WHITE}{gray}{1}
   \definecolor{RED}{rgb}{1,0,0}
   \definecolor{GREEN}{rgb}{0,1,0}
   \definecolor{BLUE}{rgb}{0,0,1}
   \definecolor{CYAN}{cmyk}{1,0,0,0}
   \definecolor{MAGENTA}{cmyk}{0,1,0,0}
   \definecolor{YELLOW}{cmyk}{0,0,1,0}
 }

\usepackage{color}
\usepackage{graphicx}
\usepackage{bm}
\usepackage{times}

\makeatother

\usepackage{babel}
\begin{document}

\title{Breaking and restoration of rotational symmetry on the  lattice for bound state multiplets}

\author{Bing-Nan Lu}
\affiliation{Institute~for~Advanced~Simulation, Institut~f\"{u}r~Kernphysik, and
J\"{u}lich~Center~for~Hadron~Physics,~Forschungszentrum~J\"{u}lich,
D-52425~J\"{u}lich, Germany}

\author{Timo~A.~L\"{a}hde}
\affiliation{Institute~for~Advanced~Simulation, Institut~f\"{u}r~Kernphysik, and
J\"{u}lich~Center~for~Hadron~Physics,~Forschungszentrum~J\"{u}lich,
D-52425~J\"{u}lich, Germany}

\author{Dean Lee}
\affiliation{Department~of~Physics, North~Carolina~State~University, Raleigh, 
NC~27695, USA}

\author{Ulf-G. Mei{\ss}ner }%
\affiliation{Institute~for~Advanced~Simulation, Institut~f\"{u}r~Kernphysik, and
J\"{u}lich~Center~for~Hadron~Physics,~Forschungszentrum~J\"{u}lich,
D-52425~J\"{u}lich, Germany}
\affiliation{Helmholtz-Institut f\"ur Strahlen- und
             Kernphysik and Bethe Center for Theoretical Physics, \\
             Universit\"at Bonn,  D-53115 Bonn, Germany}
\affiliation{JARA~-~High~Performance~Computing, Forschungszentrum~J\"{u}lich, 
D-52425 J\"{u}lich,~Germany}


\begin{abstract}
We explore the breaking of rotational symmetry on the lattice for bound state energies and practical methods for suppressing this breaking.
We demonstrate the general problems associated with lattice discretization errors and finite-volume errors using an $\alpha$ cluster model 
  for $^8$Be and $^{12}$C.
We consider the two and three $\alpha$-particle systems and focus on the lowest states with non-zero angular momentum which split into 
  multiplets corresponding to different irreducible representations of the cubic group. 
We examine the dependence of such splittings on the lattice spacing and box size.
We find that lattice spacing errors are closely related to the commensurability of the lattice with the intrinsic length scales of the system.
We also show that rotational symmetry breaking effects can be significantly reduced by using improved lattice actions, and that the physical
  energy levels are accurately reproduced by the weighted average of a given spin multiplets.
\end{abstract}


\maketitle

\section{Introduction}

In recent years, lattice Monte Carlo calculations have become a powerful tool in the study of quantum few-body and many-body systems, as well as
  relativistic field theories~\cite{Lee2009_PPNP63-117,Bazavov2010_RMP82-1349,Beane2011_PPNP66-1}.
For example, chiral effective field theory ($\chi$EFT) combined with lattice techniques has been employed to study the spectrum and structure of light 
nuclei~\cite{Borasoy2007_EPJA31-105, Epelbaum2010_EPJA45-335,Epelbaum2010_PRL104-142501,Epelbaum2011_PRL106-192501,Epelbaum2013_PRL110-112502}.
In such calculations, continuous space-time is discretized and compactified so that the path integrals can be computed numerically.
The mesh points uniformly span a cubic box and some boundary conditions such as periodic ones are imposed in each dimension. 
However, the calculated bound state energies will in general deviate from their continuum infinite-volume values due to discretization and 
  finite-volume artifacts. 

On the lattice, the full rotational symmetry group is reduced to the finite group of cubic rotations. 
Consequently, the obtained states do not unambiguously belong to a particular quantum number~\cite{Johnson1982_PLB114-147,Berg1983_NPB221-109,
  Mandula1983_NPB228-91}.
In the continuum and infinite-volume limits, quantum bound states with angular momentum $J$ form a degenerate multiplet consisting of $2J+1$ 
  members, while in lattice simulations the energy levels split into sub-groups corresponding to different irreducible representations 
  (\textit{irreps}) of the cubic group. 
The size of the energy splittings are dictated by the lattice spacing as well as the volume and boundary conditions. 

Projection onto angular momentum quantum numbers has been used to improve cubic lattice calculations of cranked Hartree-Fock~wavefuncions of 
  nuclei~\cite{Baye1984_PRC29-1056}. 
In more recent work, Dudek~\textit{et al.} have developed a method where the continuum spin of excited hadronic states in lattice QCD can be 
  reliably identified by computing overlaps with smeared lattice operators with sharply defined values of total spin~\cite{Dudek2009_PRL103-262001}.
This method has been applied to meson~\cite{Dudek2010_PRD82-034508} and baryon~\cite{Edwards2011_PRD84-074508,Meinel2012_PRD85-114510} systems. 

Davoudi~\textit{et al.} have quantified the breaking and recovery of rotational invariance as a function of the lattice spacing by means of 
  lattice operators~defined over a finite region which transform as spherical tensors with definite angular momentum in the continuum limit 
  ~\cite{Davoudi2012_PRD86-054505}. 
It was shown that such operators can be expanded in a basis of derivative operators and the corresponding operator coefficients are sensitive 
  indicators for rotational symmetry breaking.
This method has been applied perturbatively both at tree-level and the one-loop level.

The simple answer to reducing systematic errors on the lattice is to extrapolate to the limit where the lattice spacing goes to zero and
  the physical volume goes to infinity.
The infinite volume limit is rather straightforward.
The mass shifts for two-body bound states at finite volume have been studied in detail~and techniques have been developed for accelerating 
  the convergence to the infinite volume limit~\cite{Luescher1986_CMP104-177,Luescher1991_NPB354-531,Beane2004_PLB585-106,Koenig2011_PRL107-112001,
  Koenig2012_AP327-1450,Bour2011_PRD84-091503,Briceno2013_arXiv1311-7686}.
There is also numerical evidence the same techniques work for bound states with more than two constituents~\cite{Kreuzer2010_EPJA43-229,
  Kreuzer2011_PLB694-424,Koenig2011_PRL107-112001,Koenig2012_AP327-1450}.

Taking the lattice spacing to zero is a more complicated issue.
This limit is possible for a renormalizable field theory or a non-renormalizable field theory that is computed order-by-order in
  perturbation theory.
However, this approach is not suitable for non-perturbative calculations of a non-renormalizable field theory where uncancelled ultraviolet
  divergences remain.
This is the situation for many effective theories, including $\chi$EFT applied to bound states of nucleons.  

If the lattice spacing cannot be taken to zero, then the lattice improvement program proposed by Symanzik~\textit{et al.}
  ~\cite{Weisz1983_NPB212-1,Weisz1984_NPB236-397,Luescher1984_NPB240-349} provides a useful alternative approach for systematically reducing
  discretization errors.
The lattice action is systematically improved by including higher-dimensional operators to diminish the lattice spacing dependent of
  physical observables.
This method was applied to Yang-Mills theories~\cite{Weisz1983_NPB212-1,Weisz1984_NPB236-397}, gauge field theories~\cite{Curci1983_PLB130-205,
  Hamber1983_PLB133-351,Eguchi1984_NPB237-609,Luescher1984_NPB240-349} and QCD~\cite{Sheikholeslami1985_NPB259-572}. 

In this paper, we apply the technique of lattice improvement to reduce systematic errors for lattice calculations of bound state energies.
In our analysis we consider an $\alpha$ cluster model for $^8$Be and $^{12}$C on the lattice.
Our model calculations are considerably simpler than \textit{ab initio} calculations of the same $^8$Be and $^{12}$C nuclei starting from
  protons and neutrons.
However, it is sufficient to demonstrate the general issues associated with lattice discretization errors and finite-volume errors.
Furthermore, the model allows for a robust analysis over a large range of lattice spacings and cubic box sizes.
In our discussion we comment on the applications of our findings for \textit{ab initio} lattice calculations.    

In nuclear physics, clustering effects play an important role in the structure and dynamics of certain nuclei~\cite{Oertzen2006_PR432-43,
  Ebran2012_Nature487-341,Epelbaum2012_PRL109-252501,Epelbaum2013_PRL110-112502}.
For some even-even $N=Z$ nuclei, the system can be approximately described in terms of $\alpha$-clusters that interact via effective
  $\alpha$-$\alpha$ interactions.
The most frequently used interactions are local potentials involving a strongly repulsive core~\cite{Ali1966_NP80-99,Portilho1979_ZPA290-93},
  non-local ``fish-bone'' potentials~\cite{Schmid1980_ZPA297-105,Papp2008_MPLB22-2201} and other ``folding'' potentials extracted using the 
  resonating-group method~\cite{Buck1977_NPA275-246,Tang1978_PR47-167,Orabi2008_JPCS111-012045}.
In some cases a three-$\alpha$ interaction~\cite{Portilho1979_ZPA290-93} and/or forbidden state projection \cite{Tursunov2001_JPG27-1381}
  are also needed to reproduce the empirical data.  

Our objective in this paper is to study lattice errors and the energy splittings of bound state spin multiplets and practical methods 
  for reducing these artifacts.
Since our discussion is intended to be a general analysis, there is no reason to focus on some particular phenomenological interaction 
  tuned to reproduce the entire low-energy spectrum of  $^8$Be and $^{12}$C.
Instead we use a simple isotropic local potential without forbidden state projection, and we will include a three-body force to
reproduce the  ground state  energy of $^{12}$C.

\section{Theoretical framework}

\subsection{Hamiltonian}

Let $m$ denote the mass of the $\alpha$ particle.
The Hamiltonian for our multi-$\alpha$ system has the form
  \begin{equation}
  H = -\sum_{i}\frac{\nabla_{i}^{2}}{2m}+\sum_{i<j}V(|\bm{r}_{i}-\bm{r}_{j}|),\label{eq:Ham}
  \end{equation}
  where $V = V_{N}+V_{C}$ is the $\alpha-\alpha$ potential, including nuclear and Coulomb potentials.
In this paper, we consider two-$\alpha$ and three-$\alpha$ systems, corresponding to $^{8}$Be and $^{12}$C nuclei, respectively. 

\begin{figure}
\includegraphics[width=0.6 \columnwidth]{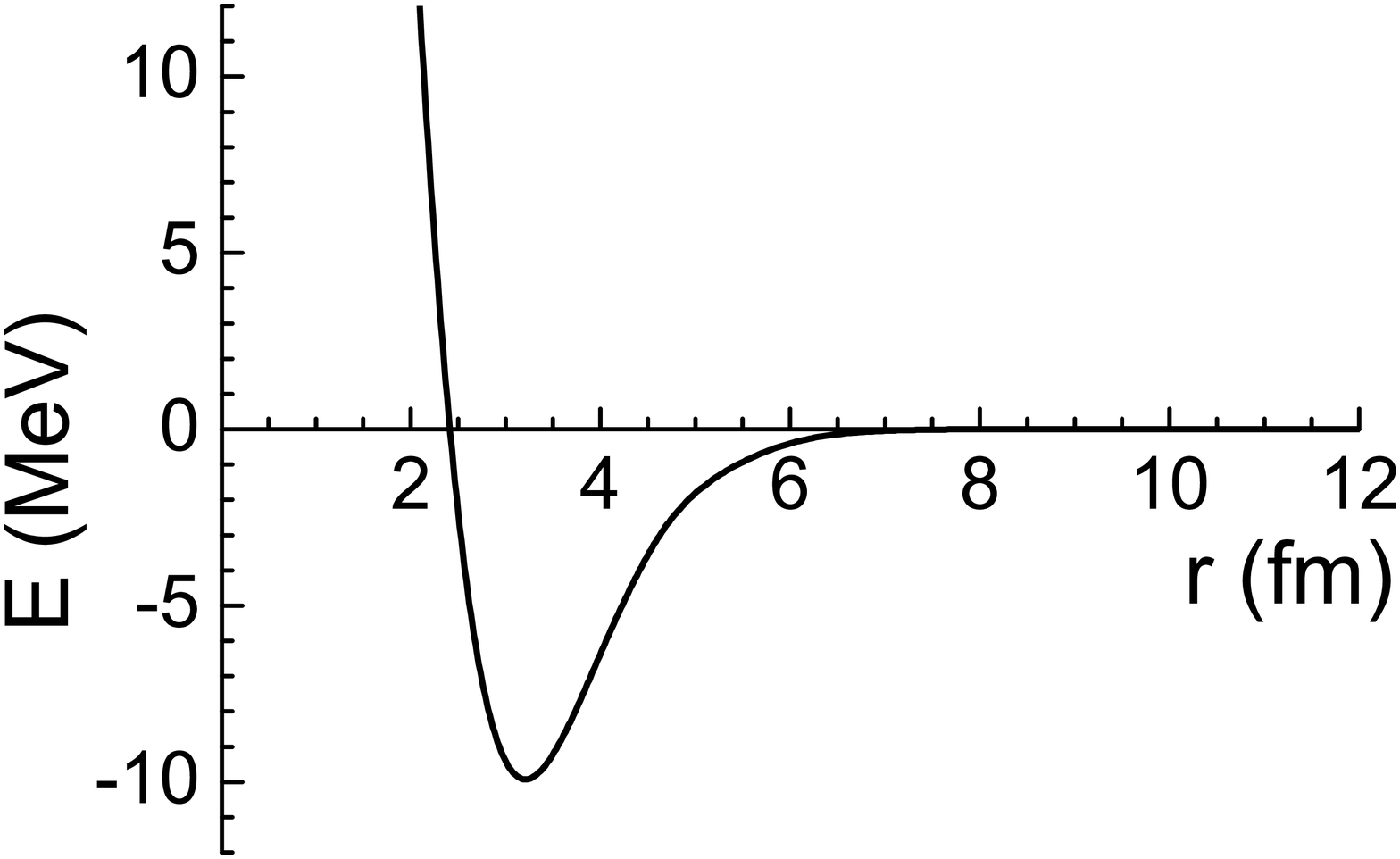}
\caption{The fitted $\alpha-\alpha$ nuclear potential as a function
of distance. \label{potential}}
\end{figure}

We use an isotropic Ali-Bodmer-type potential for the nuclear part of the $\alpha-\alpha$ interaction,
  \begin{equation}\label{VN}
  V_{N}(r)=V_{0}\,e^{-\eta_{0}^{2}r^{2}}+V_{1}\,e^{-\eta_{1}^{2}r^{2}},
  \end{equation}
  where $V_{0} = -216.346$~MeV, $V_{1} = 353.508$~MeV, $\eta_{0} = 0.436$ ~fm$^{-1}$ and $\eta_{1} = 0.529$~fm$^{-1}$.
These parameters are determined by fitting the $S$ and $D$ wave $\alpha-\alpha$ scattering lengths to their experimental
  values~\cite{Ali1966_NP80-99}.
The resulting potential is shown in Fig.~\ref{potential}.
The repulsive Coulomb potential between the $\alpha$ particles is given by
  \begin{equation}\label{VC}
  V_{C}(r)=\frac{4e^{2}}{r}{\rm erf}\left(\frac{\sqrt{3}r}{2R_{\alpha}}\right),
  \end{equation}
  where $R_{\alpha} =1.44$~fm is the radius of the $\alpha$ particle, $e$ is the unit of charge and erf denotes the error function.
We also employ the three-body interaction~\cite{Portilho1979_ZPA290-93} 
  \begin{equation}\label{3alpha}
  V(\bm{r}_{1},\bm{r}_{2},\bm{r}_{3})=V_{0}\,e^{-\lambda(r_{12}^{2}+r_{13}^{2}+r_{23}^{2})},
  \end{equation}
  where $\bm{r}_{i}$ ($i$=1,2,3) are the coordinates of the three $\alpha$ particles and $r_{12}$, $r_{13}$ and $r_{23}$ denote the 
  distances between pairs.
We take the Gaussian width parameter to be the same as in Ref.~\cite{Portilho1979_ZPA290-93}, $\lambda = 0.00506$~fm$^{-2}$, and we set 
  $V_{0} = -4.41$~MeV to recover the binding energy of the ground state of $^{12}$C.

For a system of $N$ $\alpha$ particles, we set the center of mass motion to zero and solve the Schr\"odinger equation 
  \begin{equation}
  H\,\Psi_{i}(\bm{r}_{1},\bm{r}_{2},\cdots,\bm{r}_{N-1})=E_{i}\,\Psi_{i}(\bm{r}_{1},\bm{r}_{2},\cdots,\bm{r}_{N-1})\label{eq:schroedinger}
  \end{equation}
  for the remaining $3(N-1)$ relative coordinates.
In lattice calculations the spatial vectors $\bm{r}_{1}$, $\bm{r}_{2}$ {\it etc.} assume discrete values, and Eq.~(\ref{eq:schroedinger})
  becomes a matrix eigenvalue equation.
Periodic boundary conditions for a box of size $L$ are imposed on the wave function,%
  \begin{equation}
  \psi(x+L)=\psi(x),
  \end{equation}
  where $x$ is any relative coordinate in Eq.~(\ref{eq:schroedinger}).
The Schr\"odinger equation can then be solved by diagonalization of a Hamiltonian matrix of dimension $L^{3(N-1)}\times L^{3(N-1)}$.

\subsection{Discretization of derivatives}

The spatial derivatives in Eq.~(\ref{eq:Ham}) can be expressed on the lattice by means of finite differences. 
For the second derivative we can write
  \begin{equation}
  f^{\prime\prime}(x)\approx c_{0}^{(N)}f(x)+\sum_{k=1}^{N}c_{k}^{(N)}\left[f(x+ka)+f(x-ka)\right],
  \end{equation}
  where $a$ is the lattice spacing and $c_{k}^{(N)}$ is a set of coefficients to be described in the following.
Assuming that the wave function is differentiable up to $\mathcal{O}(N)$, we can use a formula of $\mathcal{O}(N)$ to obtain a numerical
  derivative with truncation error $\mathcal{O}(a^{2N})$.
For this purpose, the coefficients $c_{k}^{(N)}$ should satisfy the following Vandermonde matrix constraint,
  \begin{equation}
  \left(\begin{array}{cccccc}
  1 & 2^{2} & 3^{2} & 4^{2} & \cdots & N^{2}_{} \vspace{.1cm} \\
  1 & 2^{4} & 3^{4} & 4^{4} & \cdots & N^{4}_{} \vspace{.1cm} \\
  1 & 2^{6} & 3^{6} & 4^{6} & \cdots & N^{6}_{} \vspace{.1cm} \\
  1 & 2^{8} & 3^{8} & 4^{8} & \cdots & N^{8}_{} \vspace{.00cm} \\
  \vdots & \vdots & \vdots & \vdots & \ddots & \vdots \vspace{.1cm} \\
  1 & 2^{2N} & 3^{2N} & 4^{2N} & \cdots & N^{2N}
  \end{array}\right)\left(\begin{array}{c}
  c_{1}^{(N)} \vspace{.05cm} \\
  c_{2}^{(N)} \vspace{.05cm} \\
  c_{3}^{(N)} \vspace{.05cm} \\
  c_{4}^{(N)} \vspace{-.05cm} \\
  \vdots \\
  c_{N}^{(N)}
  \end{array}\right)=\left(\begin{array}{c}
  a^{-2} \vspace{.15cm} \\
  0 \vspace{.15cm} \\
  0 \vspace{.15cm} \\
  0 \vspace{.0cm} \\
  \vdots \vspace{.15cm} \\
  0
  \end{array}\right).
  \end{equation}
The solution for any $N$ and $k$ is
  \begin{equation}
  c_{k}^{(N)}=a^{-2}(-1)^{k+1}\frac{2(N!)^{2}}{k^{2}(N+k)!(N-k)!},\label{eq:a_n^k-1}
  \end{equation}
  which gives
  \begin{eqnarray}
   &  & f^{\prime\prime}(x)+O(a^{2N}),\nonumber \\
   & = & c_{0}^{(N)}f(x)+a^{-2}\sum_{k=1}^{N}\frac{2(N!)^{2}(-1)^{k+1}}{k^{2}(N+k)!(N-k)!}\left[f(x+ka)+f(x-ka)\right],\nonumber \\
  \end{eqnarray}
  where $c_{0}^{(N)}$ is taken to be
  \begin{equation}
  c_{0}^{(N)}=-2\sum_{k=1}^{N}c_{k}^{(N)}.
  \end{equation}
such that $f(x)$ drops out of the final result.
Some examples of these approximations for $f^{\prime\prime}(x)$ are:
  \begin{itemize}
  \item $N=1$, 
  \begin{equation}
  \frac{1}{a^{2}}f_a(x)-\frac{2}{a^{2}}f(x),
  \end{equation}
  \item $N=2$,
  \begin{equation}
  -\frac{1}{12a^{2}}f_{2a}(x)+\frac{4}{3a^{2}}f_a(x)-\frac{5}{2a^{2}}f(x),
  \end{equation}
  \item $N=3$,
  \begin{equation}
  \frac{1}{90a^{2}}f_{3a}(x)-\frac{3}{20a^{2}}f_{2a}(x)+\frac{3}{2a^{2}}f_a(x)-\frac{49}{18a^{2}}f(x),
  \end{equation}
  \item $N=4$,
  \begin{equation}
  -\frac{1}{560a^{2}}f_{4a}(x)+\frac{8}{315a^{2}}f_{3a}(x)-\frac{1}{5a^{2}}f_{2a}(x)+\frac{8}{5a^{2}}f_a(x)-\frac{205}{72a^{2}}f(x),\label{eq:trunceq}
  \label{N=4}\end{equation}
  \end{itemize}
  where $f_{na}(x)\equiv f(x-na)+f(x+na)$.

\subsection{Broken rotational invariance}

The Hamiltonian of Eq.~(\ref{eq:Ham}) is invariant under the spatial rotational group SO(3). 
However, because the space of lattice points is only invariant under the cubic group $O$, any calculated $2J+1$ multiplet of energy
  levels is split into sub-groups belonging to different \textit{irreps} of the cubic group.
The number of levels in each representation can be determined by a group-theoretical analysis. 
\begin{table}[h]
  \caption{\label{tab:The-characters-of} 
  Characters of the classes
  of the $O$ group in different representations. $A_{1}$ through $T_{2}$ denote
  the \textit{irreps} of the $O$ group, and $D^{j}$ with $j$
  integer is the reducible cubic representation induced by the $2j+1$ dimensional SO(3) representation. 
  The dimensionality of each \textit{irreps} is given by $g$.
   $C_4({\pi}/{2})$ and $C_4^2(\pi)$ denote the class of rotations around the four-fold axis by $\pi/2$ and $\pi$, respectively.
  $C_3^\prime$ and $C_2^{\prime \prime}$ denote the class of rotations around the three- and two-fold axis, respectively.
  The dimensionality of each class is shown before the corresponding symbol. 
  }
  \begin{center}
  \begin{tabular}{cc>{\centering}p{0.4cm}ccccccccc}
  \hline 
  $D(O)$ & $g$ &  & $E$ &  & $3C_{4}^{2}$($\pi$) &  & $8C_{3}^{\prime}$ &  & $6C_{4}$($\frac{\pi}{2}$) &  & $6C_{2}^{\prime\prime}$\tabularnewline
  \cline{1-2} \cline{4-12} 
  $A_{1}$ & 1 &  & 1 &  & 1 &  & 1 &  & 1 &  & 1\tabularnewline
  $A_{2}$ & 1 &  & 1 &  & 1 &  & 1 &  & -1 &  & -1\tabularnewline
  $E$ & 2 &  & 2 &  & 2 &  & -1 &  & 0 &  & 0\tabularnewline
  $T_{1}$ & 3 &  & 3 &  & -1 &  & 0 &  & 1 &  & -1\tabularnewline
  $T_{2}$ & 3 &  & 3 &  & -1 &  & 0 &  & -1 &  & 1\tabularnewline
  $D^{J}$ & $2J+1$ &  & $2J+1$ &  & $(-1)^{J}$ &  & $1-{\rm mod}(J,3)$ &  & $(-1)^{[\frac{J}{2}]}$ &  & $(-1)^{J}$\tabularnewline
  
  \hline 
  \end{tabular}
  \end{center}
\end{table}

In Table~\ref{tab:The-characters-of}, we list the characters of the $O$ representations. 
The decomposition of the SO(3) induced representation can be obtained by calculating the inner product of the character vectors
  in group space. 
The results for angular momenta  up to $J=8$ are shown in Table~\ref{tab:The-coefficients-for}.
The splitting of each $2J+1$ multiplet can be read from the corresponding column.
For example, the three levels from $J=1$ multiplet do not split, while the five levels from the $J=2$ multiplet assume a $2+3$ structure.

In order to obtain the transformtion matrices, we must specify the individual wave functions with good quantum numbers.
In lattice calculations the angular momenta are no longer exactly conserved.
Nevertheless, we can define the quantum number $J_z$ through the relation
  \begin{equation}
    R_z(\frac{\pi}{2}) = \exp \left( -i\frac{\pi}{2}J_z \right),
  \end{equation}
  where $R_z(\pi / 2)$ is a rotation around the $z$ axis by $\pi / 2$ which is an element of $O$ group.
In this case $J_z$ are integers modulo 4.
Since the eigenvalues of $R_z(\pi / 2)$ are not degenerate in any \textit{irreps} of the cubic group,
the levels in each representation can be distinguished unambiguously according to their respective $J_z$ quantum numbers.

\begin{table}[h]
  \caption{\label{tab:The-coefficients-for}
  Coefficients for decomposing the induced representations $D^{J}$ in terms of the irreducible representations of the $O$ group.
  The dimensionality of each irreducible representation is given by $g$.
  }
  \begin{center}
  \begin{tabular}{ccc|cccccccccc}
  $D(O)$ & $g$ &  &  & $D^{0}$ & $D^{1}$ & $D^{2}$ & $D^{3}$ & $D^{4}$ & $D^{5}$ & $D^{6}$ & $D^{7}$ & $D^{8}$\tabularnewline
  \hline 
  $A_{1}$ & 1 &  &  & 1 & 0 & 0 & 0 & 1 & 0 & 1 & 0 & 1\tabularnewline
  $A_{2}$ & 1 &  &  & 0 & 0 & 0 & 1 & 0 & 0 & 1 & 1 & 0\tabularnewline
  $E$ & 2 &  &  & 0 & 0 & 1 & 0 & 1 & 1 & 1 & 1 & 2\tabularnewline
  $T_{1}$ & 3 &  &  & 0 & 1 & 0 & 1 & 1 & 2 & 1 & 2 & 2\tabularnewline
  $T_{2}$ & 3 &  &  & 0 & 0 & 1 & 1 & 1 & 1 & 2 & 2 & 2\tabularnewline
  \end{tabular}
  \end{center}
\end{table}

In Eq.~(\ref{eq:decomposition}) we show the decompositions of the first few ($J\leq 3$) \textit{irreps} of the SO(3) group.
Here we employ the spherical harmonics $Y_{l,m}$, with $l$, $m$ integers, as the basis.
In each bracket the basis spanning the representation of the $O$ group is listed.
Note that only the functions with the same $J_z$ (mod 4) are mixed.

\begin{eqnarray}
	\mathcal{H}_{J=0}= &  & A_{1}\left[Y_{0,0}\right],\nonumber \\
	\mathcal{H}_{J=1}= &  & T_{1}\left[Y_{1,0},\; Y_{1,\pm1}\right],\nonumber \\
	\mathcal{H}_{J=2}= &  & E\left[\sqrt{\frac{1}{2}}Y_{2,2}+\sqrt{\frac{1}{2}}Y_{2,-2},\; Y_{2,0}\right]\nonumber \\
	 & \oplus & T_{2}\left[\sqrt{\frac{1}{2}}Y_{2,2}-\sqrt{\frac{1}{2}}Y_{2,-2},\; Y_{2,\pm1}\right],\nonumber\\
	\mathcal{H}_{J=3}= &  & A_{2}\left[\sqrt{\frac{1}{2}}Y_{3,2}-\sqrt{\frac{1}{2}}Y_{3,-2}\right]\nonumber \\
	 & \oplus & T_{1}\left[\sqrt{\frac{5}{8}}Y_{3,\mp3}+\sqrt{\frac{3}{8}}Y_{3,\pm1},\; Y_{3,0}\right]\nonumber \\
	 & \oplus & T_{2}\left[\sqrt{\frac{5}{8}}Y_{3,\pm1}-\sqrt{\frac{3}{8}}Y_{3,\mp3},\;\sqrt{\frac{1}{2}}Y_{3,2}+\sqrt{\frac{1}{2}}Y_{3,-2}\right].\nonumber\\
\label{eq:decomposition}
\end{eqnarray}

\section{Results and discussion}

\subsection{The $^{8}$Be nucleus}

We note that the $^{8}$Be nucleus is not bound.
However, for  the purposes of this analysis on bound state lattice calculations, we can artificially produce shallow $^{8}$Be bound 
  states by increasing $V_{0}$ by an amount 30\% larger in magnitude than the value $V_{0} = -4.41$~MeV we quoted earlier.
With this strengthened potential, the $^{8}$Be nucleus has a ground state at $E(0^{+})=-10.8$~MeV and one excited state at $E(2^{+})=-3.3$~MeV. 
These energies are measured relative to the $\alpha$-$\alpha$ threshold. 
On the lattice, the $2^+$ state splits into two multiplets corresponding to the $E$ and $T_{2}$ representations of the $O$ group.
These splittings arise due to finite volume as well as lattice dicretization effects.

\begin{figure}[b]
  \begin{center}
  \includegraphics[width=0.7\columnwidth]{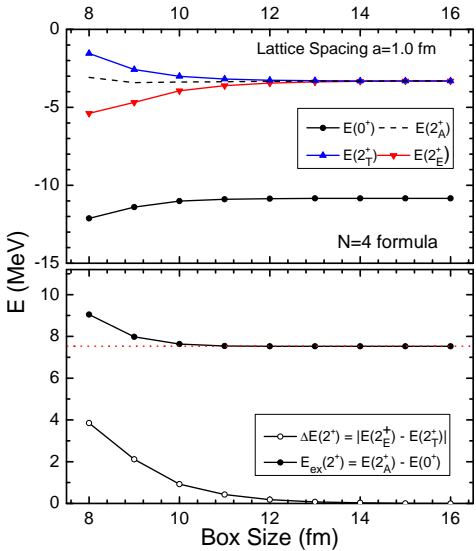}
  \caption{\textit{\small \label{fig:two-alpha-boxsize}Upper panel}{\small :
  Lowest $0^{+}$ and $2^{+}$ levels of $^{8}$Be nucleus versus box size. The lattice spacing $a$ is fixed at 1.0 fm with lattice improvement index $N=4$.
  The  $2^{+}$ states split into two multiplets $2_{{\rm E}}^{+}$(down
  triangles) and $2_{{\rm T}}^{+}$(up triangles) corresponding to different
  representations of the $O$ group. The dashed line represents the
  weighted averaged value $E(2_{{\rm A}}^{+})$.
  }\textit{\small Lower panel}{\small : The energy splitting $\Delta E(2^+)$ (open circles) and the average excitation
  gap $E_{{\rm ex}}(2^{+})$ (full circles)  versus box size. The dotted line shows the continuum infinite-volume limit result for $E_{{\rm ex}}(2^{+})$.}}
  \end{center}
\end{figure}

\begin{figure}[t]
  \begin{center}
  \includegraphics[width=0.7\columnwidth]{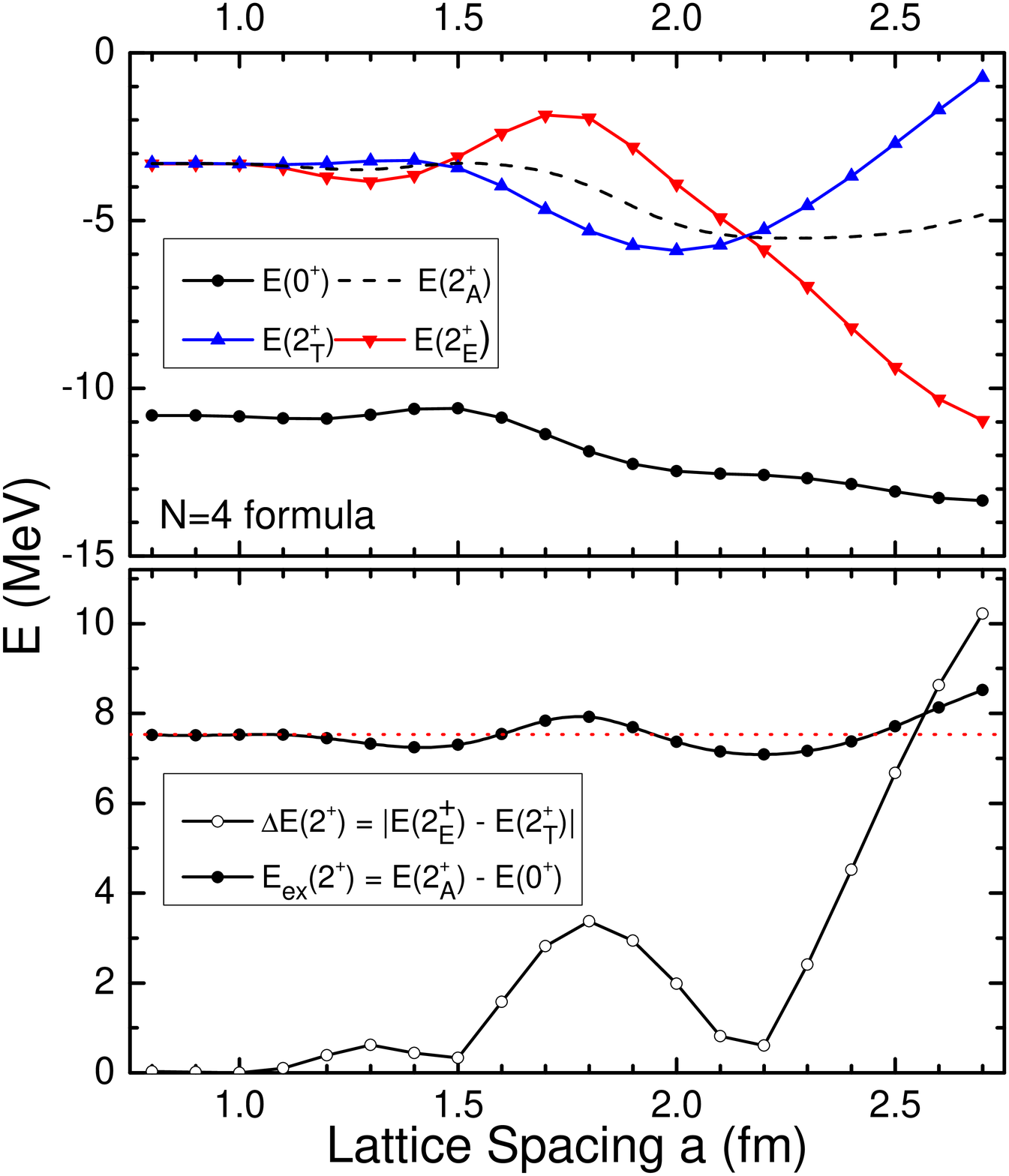}
  \caption{{\small \label{fig:The-same-as}\textit{Upper panel}{\small :
  Lowest $0^{+}$ and $2^{+}$ levels of $^{8}$Be nucleus versus lattice spacing with lattice improvement index $N=4$. 
  The box size $L$ is kept larger than 15 fm to remove finite volume effects.
  The dashed line represents the weighted averaged value $E(2_{{\rm A}}^{+})$.
  }\textit{\small Lower panel}{\small : The energy splitting $\Delta E(2^+)$ (open circles) and the average excitation gap
    $E_{{\rm ex}}(2^{+})$ (full circles) versus lattice spacing.
  The dotted line shows the continuum infinite-volume limit result for $E_{{\rm ex}}(2^{+})$.}}}
  \end{center}
\end{figure}

\begin{figure}[b]
  \includegraphics[width=\columnwidth]{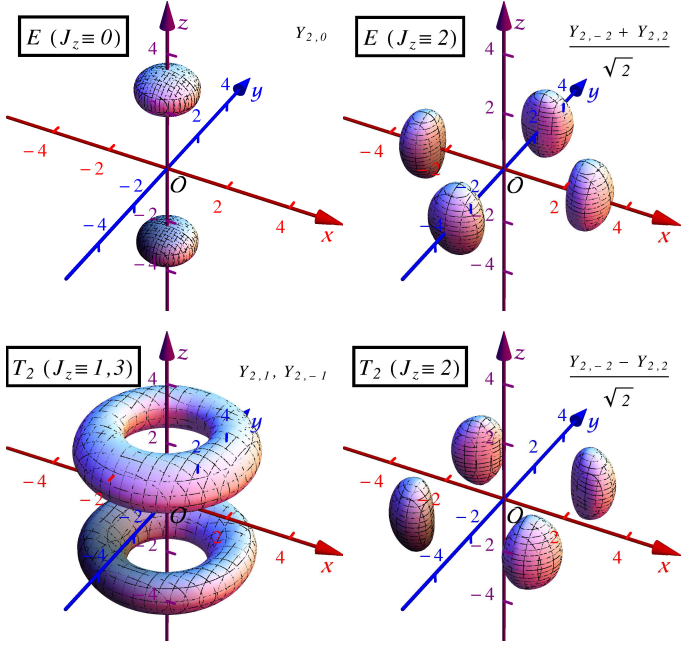}
  \caption{{\small \label{fig:The-probability-density}
    The 3-d probability density distributions $|\psi(\bm{r})|^2$ of the $\alpha$-$\alpha$ separation in the $^{8}$Be nucleus
   	  calculated with lattice spacing $a=0.6$ fm and derivative improving index $N=4$.
    In each sub-figure we show the isohypsic surface with 70\% of the maximal probability density.
  	The coordinates on $x$, $y$ and $z$ axes are in fm.
  	\textit{Upper Panels:}{\small We show the $2_{\rm E}^{+}$ state with $J_z\equiv 0$ and $J_z\equiv 2$, respectively.}
  	\textit{Lower Panels:}{\small We show the $2_{\rm T}^{+}$ state with $J_z\equiv 1,3$ and $J_z\equiv 2$, respectively.}
  	Note that the probability density distributions of the $2_{\rm T}^{+}$ state with $J_z\equiv 1$ ($\psi(\bm{r}) \propto Y_{2,1}$)  
  	  and $J_z\equiv 3$ ($\psi(\bm{r}) \propto Y_{2,-1}$) are exactly the same due to time-reversal symmetry, see Eq.~(\ref{eq:decomposition}).
  }}
\end{figure}

We start with a small lattice spacing, $a = 1.0$~fm, where the lattice discretization errors should be small.
The derivative improvement index is chosen to be $N=4$, corresponding with Eq.~(\ref{N=4}). 
In the upper panel of Fig.~\ref{fig:two-alpha-boxsize}, we show the calculated lowest $0^+$ and $2^+$ energy levels as functions of box size $L$.
For small $L$, the $2_{{\rm T}}^{+}$ energy is pushed upward, while the $2_{\rm E}^{+}$ energy is pushed downward.
The splitting between these two energies is shown explicitly in the lower panel of Fig.~\ref{fig:two-alpha-boxsize}.
Our results are consistent with the finite-volume energy shift formulas derived in Ref.~\cite{Koenig2011_PRL107-112001,Koenig2012_AP327-1450}. 
As we increase the box size, these levels merge for $L\geq15$ fm and are consistent with the continuum infinite-volume result. 
This matches our expectation that lattice discretization errors are negligible at this small lattice spacing and the splittings are due only to 
  finite volume effects.

We now define a multiplet-averaged energy for the $2^+$ state,  
\begin{equation}
  E(2_{{\rm A}}^{+}) = (2E(2_{{\rm E}}^{+})+3E(2_{{\rm T}}^{+}))/5.
\end{equation}
This is shown as a dashed line in the upper panel of Fig.~\ref{fig:two-alpha-boxsize}.
The weight factors 2 and 3 denote the number of members for each cubic representation.
This multiplet-averaged energy appears much closer to continuum infinite-volume result than either of the two individual branches, 
  $E(2_{{\rm E}}^{+})$ and $E(2_{{\rm T}}^{+})$.
Our definition for $E(2_{{\rm A}}^{+})$ is motivated by the fact that averaging over all elements of spin multiplets simplify the $L$ dependence
  of the finite volume corrections for two-body bound states~\cite{Koenig2011_PRL107-112001,Koenig2012_AP327-1450}. 
In fact, the multiplet-averaged finite-volume energy correction has a simple universal form, where the sign of the correction alternates for
  even and odd $L$.

In some cases it is also useful to consider the multiplet-averaged excitation gap, 
  \begin{equation}
   E_{{\rm ex}}(2^{+}) = E(2_{{\rm A}}^{+})-E(0^{+}),
  \end{equation}
  to help cancel similar errors in the $0^{+}$ and $2^{+}$ levels due to the finite box size. 
In the lower panel of Fig.~\ref{fig:two-alpha-boxsize}, these results are depicted by filled circles.
We see that $E_{{\rm ex}}(2^{+})$ converges faster to its infinite volume limit than the energy splitting, 
  \begin{equation}
   \Delta E(2^+) = |E(2_{{\rm E}}^{+})-E(2_{{\rm T}}^{+})|. 
  \end{equation}
We have used the absolute value in this definition of the splitting simply for convenience in plotting.
A box size of $L=10$~fm is large enough for an estimation of the excitation energy $E_{{\rm ex}}(2^{+})$ with error less than 100~keV, 
  while the $\Delta E(2^+)$ energy splitting still exceeds 1~MeV.

\begin{figure}[t]
  \includegraphics[width=0.49\columnwidth]{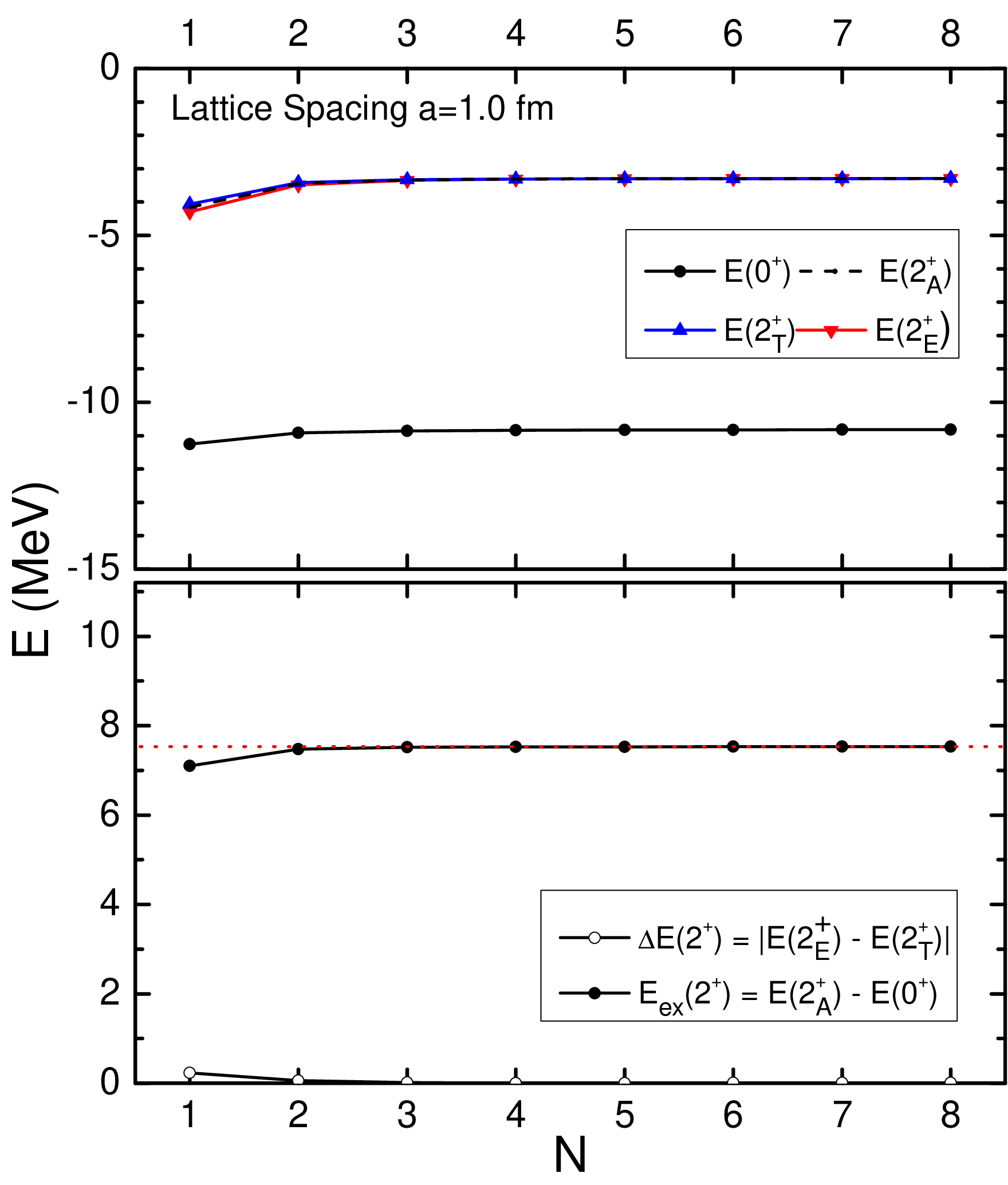}
  \includegraphics[width=0.49\columnwidth]{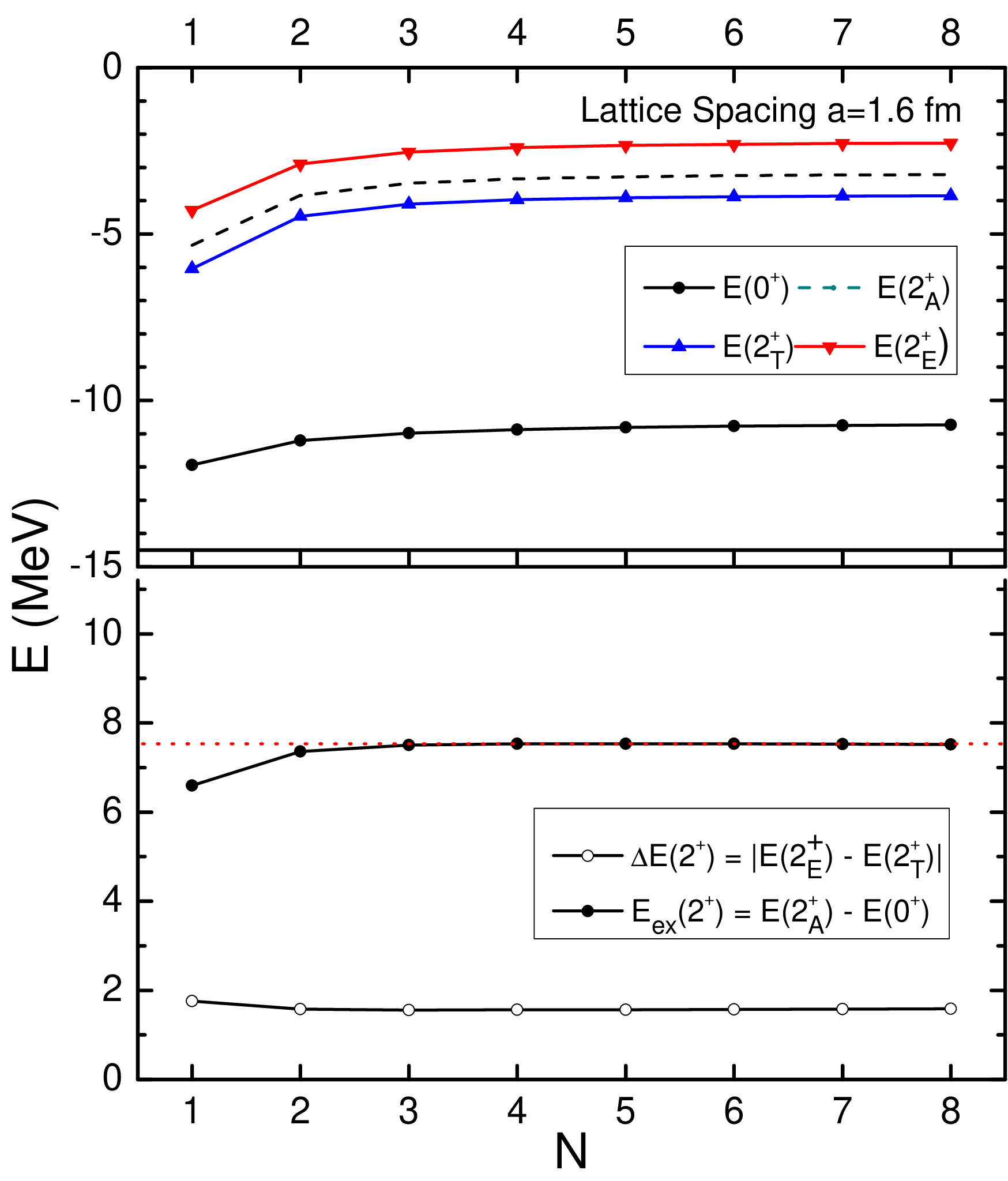}
  \includegraphics[width=0.49\columnwidth]{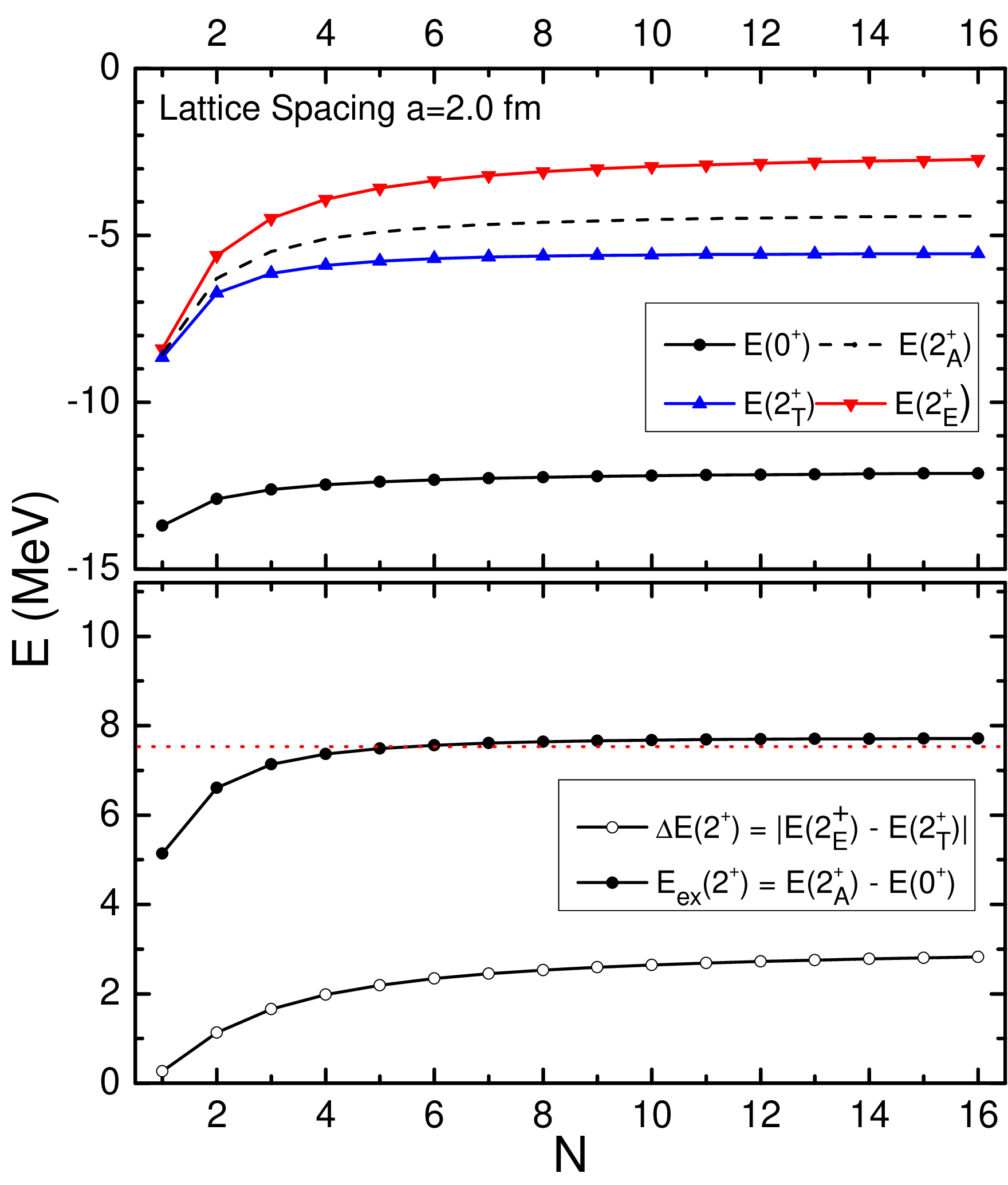}
  \includegraphics[width=0.49\columnwidth]{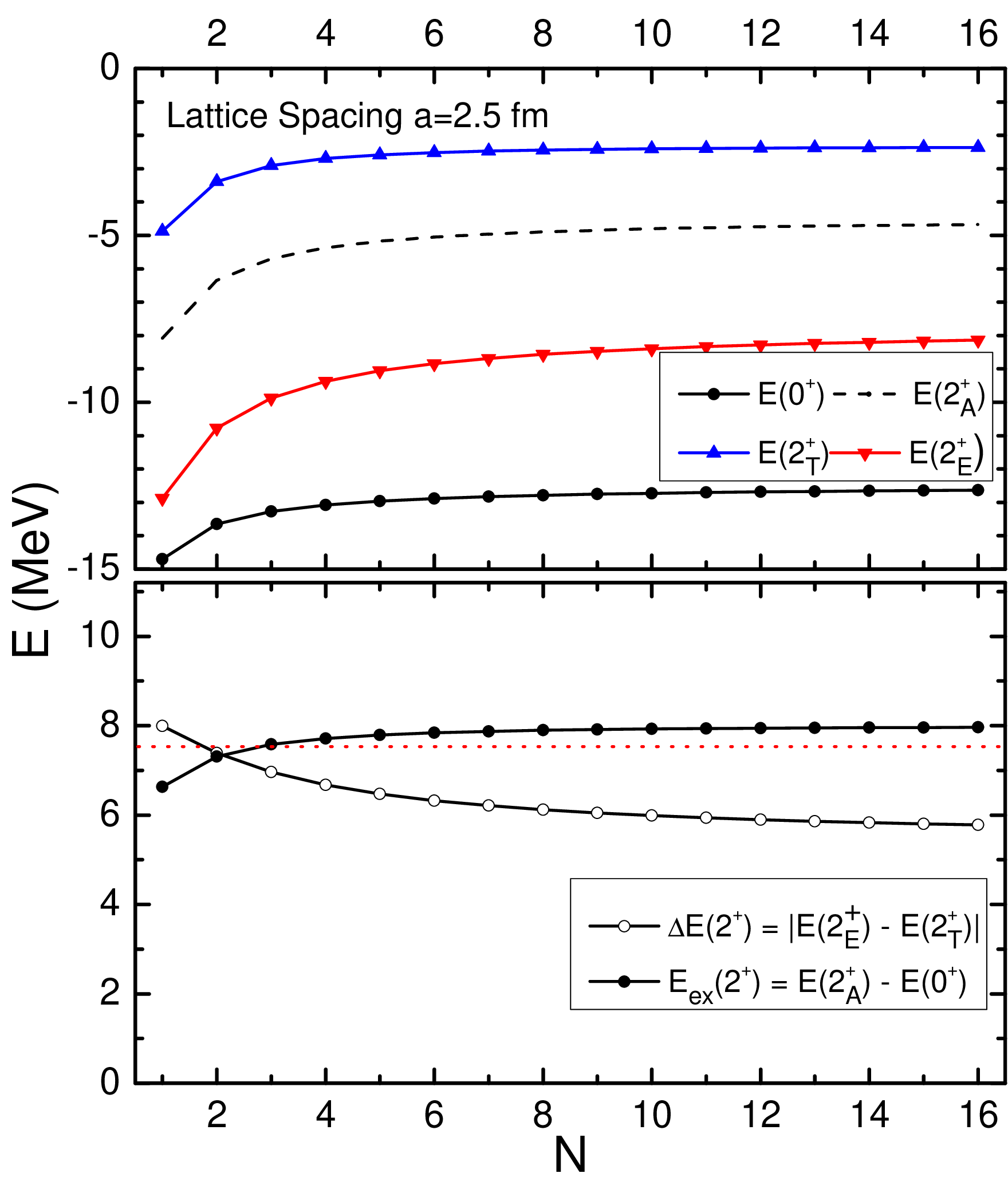}
  \caption{{\small \label{fig:The-same-as-1}
    \textit{Upper panels of each subfigure}{\small :
    Lowest $0^{+}$ and $2^{+}$ levels of $^{8}$Be nucleus versus lattice improvement index $N$ for lattice spacings
	  $a = 1.0, 1.6, 2.0, 2.5~{\rm fm}$. 
	The box size $L$ is kept larger than 15 fm to remove finite volume effects.  
    The dashed lines represents the weighted averaged value $E(2_{{\rm A}}^{+})$.
    }\textit{\small Lower panels of each subfigure}{\small : 
	The energy splitting $\Delta E(2^+)$ (open circles) and the average excitation gap $E_{{\rm ex}}(2^{+})$ (full circles)  
	  versus lattice improvement index $N$ for lattice spacings $a = 1.0, 1.6, 2.0, 2.5~{\rm fm.}$
	The dotted line shows the continuum infinite-volume limit result for $E_{{\rm ex}}(2^{+})$.  }}}
\end{figure}

Next, let us turn our attention to systematic errors due to nonzero lattice spacing.
In Fig.~\ref{fig:The-same-as}, we show the $0^{+}$ and $2^{+}$ levels, $E(2_{{\rm A}}^{+})$, $E_{{\rm ex}}(2^{+})$, and $\Delta E(2^+)$ as
  a function of the lattice spacing $a$. 
In all cases the box size $L$ is taken to be $L\geq15$~fm and the derivative improvement index is $N=4$.
For $a\leq1.0$~fm, the two branches $E(2_{{\rm E}}^{+})$ and $E(2_{{\rm T}}^{+})$ merge, and for larger $a$ they split apart and show 
  some oscillations.
For $a\geq2.5$~fm the splittings are as large as 10~MeV.
 
Before discussing the physics behind the oscillatory behavior of the energy splitting, we note that although $\Delta E(2^+)$ 
  becomes as large as several MeV, the error of the multiplet-averaged excitation gap $E_{{\rm ex}}(2^{+})$ shown in the lower panel
  of Fig.~\ref{fig:The-same-as} does not exceed $0.5$~MeV for $a\leq 2.5$~fm. We note that averaging over the $2^{+}$ mutliplet and
  the cancellation of errors between the $0^{+}$ and $2^{+}$ levels both play important roles in attaining this accuracy.
This technique of calculating multiplet-averaged excitation gaps would seem to be a useful tool for reducing systematic errors
  in any lattice calculation of bound state energies.
 
The energy splitting $\Delta E(2^+)$ in the lower panel of Fig.~\ref{fig:The-same-as} is not a monotonic function of $a$. 
We can observe two zeros near $a=1.5$~fm and $2.2$~fm where the two branches $E(2_{{\rm E}}^{+})$ and $E(2_{{\rm T}}^{+})$ cross.
For $a\leq1.0$~fm the splitting is negligible. For $1.0 \leq a\leq 1.5$~fm, the $2_{{\rm T}}^{+}$ is higher than the $2_{{\rm E}}^{+}$.
However, in the region $1.5 \leq a\leq 2.2$~fm, the order is reversed with the $2_{{\rm T}}^{+}$ is lower than the $2_{{\rm E}}^{+}$. 
Then for $a\geq 2.2$~fm, the $2_{{\rm E}}^+$ is once again lower and the splitting increases monotonically.

\begin{figure}[t]
  \begin{center}
  \includegraphics[width=0.7\columnwidth]{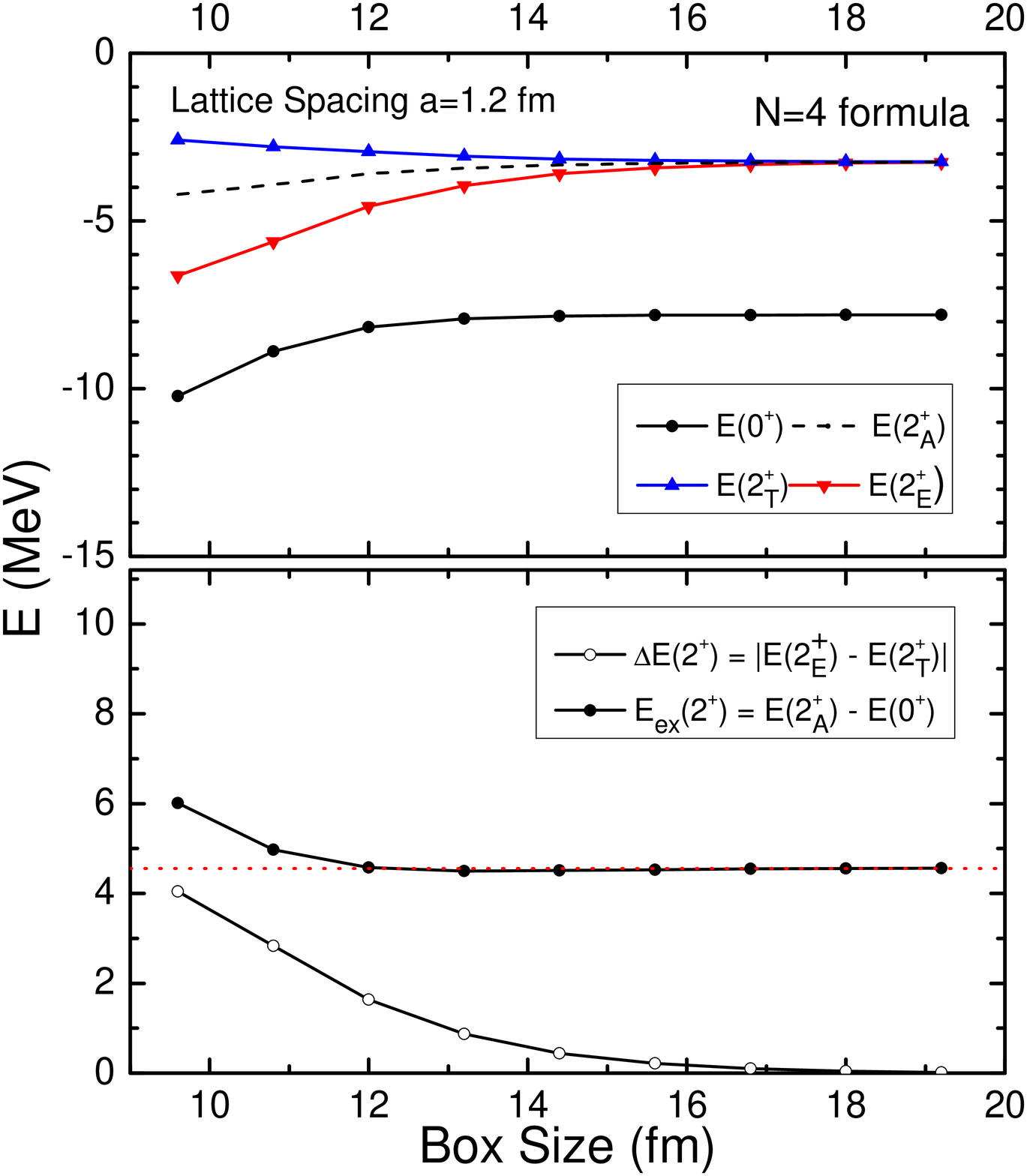}
  \caption{\textit{\small \label{fig:Upper-panel:-Lowest}Upper panel}{\small :
  Lowest $0^{+}$ and $2^{+}$ levels of $^{12}$C nucleus versus box size. The lattice spacing $a$ is fixed at 1.2 fm with lattice improvement index $N=4$.
  The  $2^{+}$ states split into two multiplets $2_{{\rm E}}^{+}$(down
  triangles) and $2_{{\rm T}}^{+}$(up triangles) corresponding to different
  representations of the $O$ group. The dashed line represents the
  weighted averaged value $E(2_{{\rm A}}^{+})$.
  }\textit{\small Lower panel}{\small : The energy splitting $\Delta E(2^+)$ (open circles) and the average excitation
  gap $E_{{\rm ex}}(2^{+})$ (full circles) versus box size. The dotted line shows the continuum infinite-volume limit result for $E_{{\rm ex}}(2^{+})$.}}
  \end{center}
\end{figure}

\begin{figure}[b]
  \begin{center}
  \includegraphics[width=0.7\columnwidth]{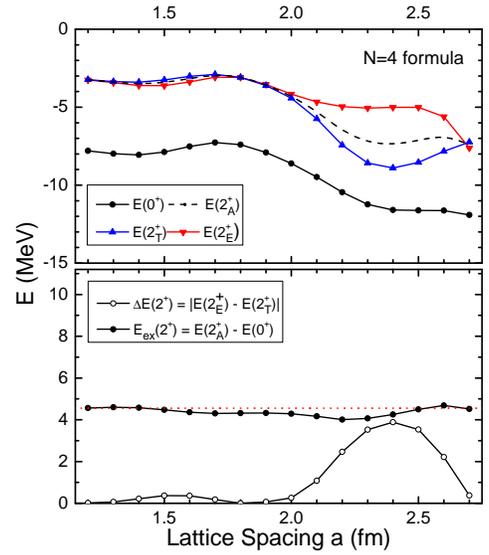}
  \caption{{\small \label{fig:12C_versus_a}\textit{Upper panel}{\small :
  Lowest $0^{+}$ and $2^{+}$ levels of $^{12}$C nucleus versus lattice spacing with lattice improvement index $N=4$. 
  The box size $L$ is kept larger than 18 fm to remove finite volume effects.
  The dashed line represents the weighted averaged value $E(2_{{\rm A}}^{+})$.
  }\textit{\small Lower panel}{\small : The energy splitting $\Delta E(2^+)$ (open circles) and the average excitation gap
    $E_{{\rm ex}}(2^{+})$ (full circles) versus lattice spacing.
  The dotted line shows the continuum infinite-volume limit result for $E_{{\rm ex}}(2^{+})$.}}}
  \end{center}
\end{figure}

These observed oscillations can be explained by examining the probability distribution as a function of the relative separation
  vector between $\alpha$ particles.
In Fig.~\ref{fig:The-probability-density} we show the calculated probability distributions for the state of
  $2^{+}_{\rm E}$ with $J_z\equiv 0,2$ and the state of $2^+_T$ with $J_z\equiv 1,3$ and $2$.
Let $\vec{e}_{x}$, $\vec{e}_{y}$ and $\vec{e}_{z}$ be the lattice unit vectors. 
For the $2_{{\rm E}}^{+}$ states, the probability distributions are peaked along the axis directions $\pm\vec{e}_{x}$, 
  $\pm\vec{e}_{y}$ (for $J_z\equiv 2$) or $\pm\vec{e}_{z}$ (for $J_z\equiv 0$).
On the other hand, for the $2_{{\rm T}}^{+}$ states, the probability distributions are either peaked along the diagonal directions
  $(\pm\vec{e}_{x}\pm\vec{e}_{y})/\sqrt{2}$ (for $J_z\equiv 2$)
  or assume a ring-like shape pointing along $(\vec{e}_{x}\cos\theta+\vec{e}_{y}\sin\theta\pm\vec{e}_{z})/\sqrt{2}$ with
  $\theta$ an arbitrary angle (for $J_z\equiv 1,3$).  

In the continuum infinite-volume limit, the average distance between the $\alpha$ particles is $d \approx 2.9~{\rm fm}$ 
  in all the $2^+$ states considered here.
Therefore we expect the $2^{+}_{\rm E}$ energy to be minimized when the average separation distance $d$ is commensurate
  with distances between lattice points along the coordinate axes.  
This corresponds with minima for lattice spacings $a = d/N$ for integer $N$, or $a \approx 2.9~{\rm fm}$, $1.45~{\rm fm}$, 
  etc., and maxima in between these values.
This is a good description of what is seen in Fig.~\ref{fig:The-same-as}. 
Similarly we expect the $2^{+}_{\rm T}$ energy to be minimized when the average separation distance $d$ is commensurate
  with distances between lattice points along the $(\pm\vec{e}_{x}\pm\vec{e}_{y})/\sqrt{2}$ ($J_z\equiv 1,2,3$) or
  $(\pm\vec{e}_{x}\pm\vec{e}_{y}\pm\vec{e}_{z})/\sqrt{3}$ ($J_z\equiv 1,3$) diagonal directions. 
This corresponds with minima for lattice spacings $a = d/(\sqrt{3}N)$ or $a = d/(\sqrt{2}N)$ for integer $N$, or 
  $a \approx 2.05~{\rm fm}$ or $1.67~{\rm fm}$, $1.03~{\rm fm}$ or $0.84~{\rm fm}$, etc., 
  and maxima in between these values.  This is also a good description of what is seen in Fig.~\ref{fig:The-same-as}.    

We now consider the usefulness of improved lattice actions in reducing discretization errors.
In this analysis we only consider improvements to the lattice dispersion relations and keep the same functional form for 
  the interaction inherited from our smooth continuum potential.
For  calculations with short-range interactions such as cutoff-dependent contact interactions, one should also consider 
  improvements to these short-range interactions. 
However, there are practical reasons for restricting our attention here to improvements of the lattice dispersion relation.
In \textit{ab initio} nuclear simulations the $\alpha$ clusters are built from protons and neutrons.
Therefore the details of the interactions between $\alpha$ clusters are difficult to compute and difficult to control via
  the underlying lattice action of the nucleons.
However, it is much easier to measure the $\alpha$ dispersion relation and even to modify the $\alpha$ dispersion relation 
  via the underlying nucleon lattice dispersion relation. This is why in this analysis we consider only improvements to
  the lattice dispersion relations. 

In Fig.~\ref{fig:The-same-as-1}, we show the $0^{+}$ and $2^{+}$ levels, $E(2_{{\rm A}}^{+})$, $E_{{\rm ex}}(2^{+})$, 
  and $\Delta E(2^+)$ as a function of the derivative improvement index $N$ for lattice spacings $a = 1.0, 1.6, 2.0, 2.5~{\rm fm}$.
In all cases the box size $L$ is taken to be $L\geq15$~fm. 
In the lower panel of each subfigure, the dotted line shows the continuum infinite-volume limit result for $E_{{\rm ex}}(2^{+})$.
For $a=1.0$~fm, the splitting $\Delta E(2^+)$ is negligible and the excitation gap $E_{{\rm ex}}(2^{+})$ is nearly indistinguishable
  from the continuum value for $N\geq 3$. 
For the case of $a=1.6$~fm, the excitation gap $E_{{\rm ex}}(2^{+})$ reaches a plateau with increasing $N$ close to the continuum result.
The splitting $\Delta E(2^+)$ also converges with increasing $N$ to a small non-zero value. 
Although the small errors in $E_{{\rm ex}}(2^{+})$ and $\Delta E(2^+)$ are not completely eliminated by the use of higher-order 
  lattice actions for the kinetic energy, the results are clearly much more reliable with the improved actions $N$.
We expect that the remaining discretization errors can be completely removed by including lattice improvements to the interaction also.

It is now interesting to observe what happens for larger lattice spacings.
For $a=2.0$~fm and $a=2.5$~fm, the splitting $\Delta E(2^+)$ in fact increases as a function of $N$.
This reflects the fact that lattice improvements to the interaction must also be included also to see systematic improvement with $N$.
However, these missing improvements to the interaction are not so important for the average excitation gap $E_{{\rm ex}}(2^{+})$.
The error of the averaged excitation energy $E_{{\rm ex}}(2^{+})$ is monotonically decreasing function of $N$ and comes very close
  to the continuum value, with an error of about $0.5$~MeV at $a=2.5~{\rm fm}$. 
This suggests that the process of calculating multiplet-averaged excitation gaps with improved dispersion actions can significantly
  decrease lattice discretization errors even at large lattice spacing.

\subsection{The $^{12}$C nucleus}

For the $^{12}$C nucleus, we will use the potentials in Eq.~(\ref{VN}) and (\ref{VC}) without any modification to $V_0$ and include
  the three-body interaction in Eq.(\ref{3alpha}) to get the physical ground-state binding energy.
We now go through the same analysis that we applied to $^8$Be. 
In the upper panel of Fig.~\ref{fig:Upper-panel:-Lowest}, we show the calculated lowest $0^+$ and $2^+$ energy levels with a small
  fixed lattice spacing of $a=1.2$~fm versus box sizes $L$.
As we found for $^8$Be, the $2^{+}$ states split into two multiplets, $2_{{\rm E}}^{+}$(down triangles) and $2_{{\rm T}}^{+}$
  (up triangles) corresponding to different representations of the $O$ group. 
Therefore we make the analogous definitions for the weighted average $E(2_{{\rm A}}^{+})$, multiplet splitting $\Delta E(2^+)$, 
  and average excitation gap $E_{{\rm ex}}(2^{+})$.

In the lower panel we show the energy splitting $\Delta E(2^+)$ and the average excitation gap $E_{{\rm ex}}(2^{+})$ versus box size.
The dotted line shows the continuum infinite-volume limit result for $E_{{\rm ex}}(2^{+})$. 
In this case, the lattice errors become inegligible for $L\geq18$~fm.
In what follows, we set the  box size to $L\geq18$~fm so that the errors are dominated by lattice discretization effects. 

In Fig.~\ref{fig:12C_versus_a}, we show the $0^{+}$ and $2^{+}$ levels of $^{12}$C, $E(2_{{\rm A}}^{+})$, $E_{{\rm ex}}(2^{+})$, 
  and $\Delta E(2^+)$ as functions of the lattice spacing $a$ for derivative improvement index $N=4$.
The results are similar to what we found for $^{8}$Be, and can be interpreted along similar lines.
As note in Ref.~\cite{Epelbaum2012_PRL109-252501}, the $\alpha$ clusters in the $0^{+}$ and $2^{+}$ states form a compact
  triangular geometry.
The oscillations in energy as a function lattice spacing are due to commensurability of the underlying lattice mesh with
  this triangular structure.
For example, at $a=1.5$~fm and $a=3.0$~fm (not shown), the $2_{{\rm E}}^{+}$ level has a minimum because in this case,
  two pairs of $\alpha-\alpha$ separations in the three-$\alpha$ system are commensurate with lattice points along the 
  coordinate axes.
We note also that average excitation gap $ E_{{\rm ex}}(2^{+})$ is suprisingly accurate even for large lattice spacings.
The error is less than 1~MeV for lattice spacings $a < 2.7$~fm.

\begin{figure}
  \includegraphics[width=0.49\columnwidth]{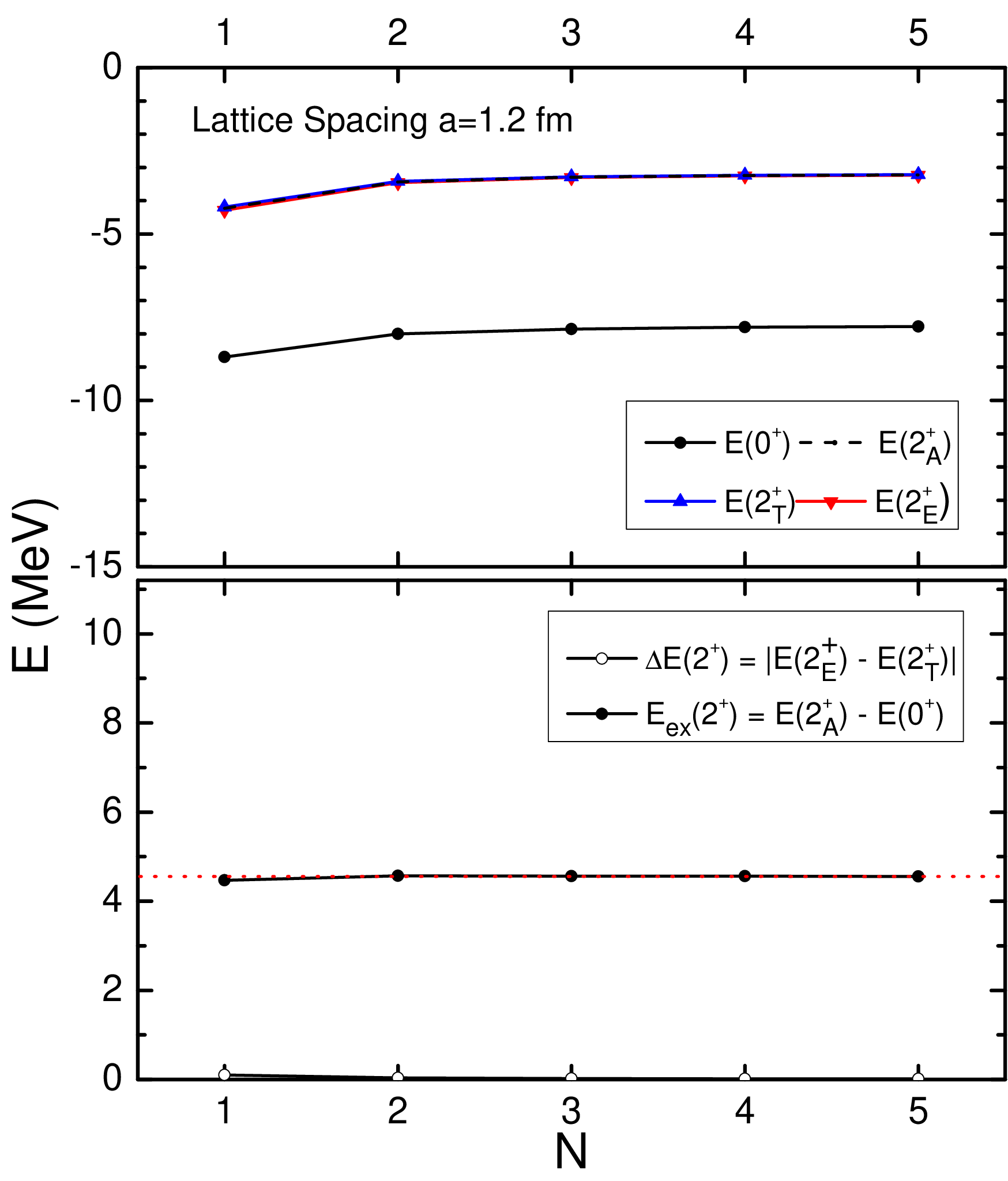}
  \includegraphics[width=0.49\columnwidth]{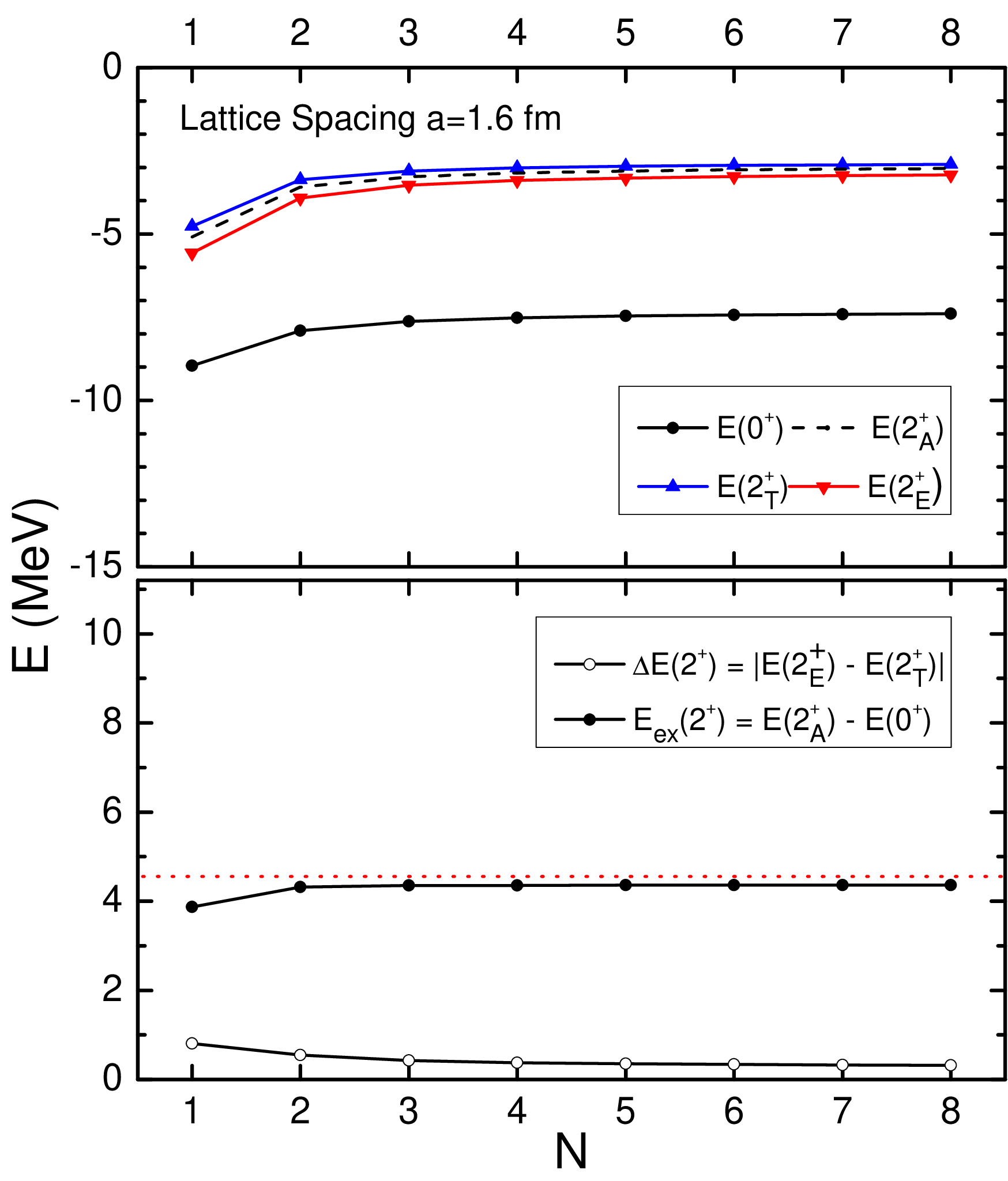}
  \includegraphics[width=0.49\columnwidth]{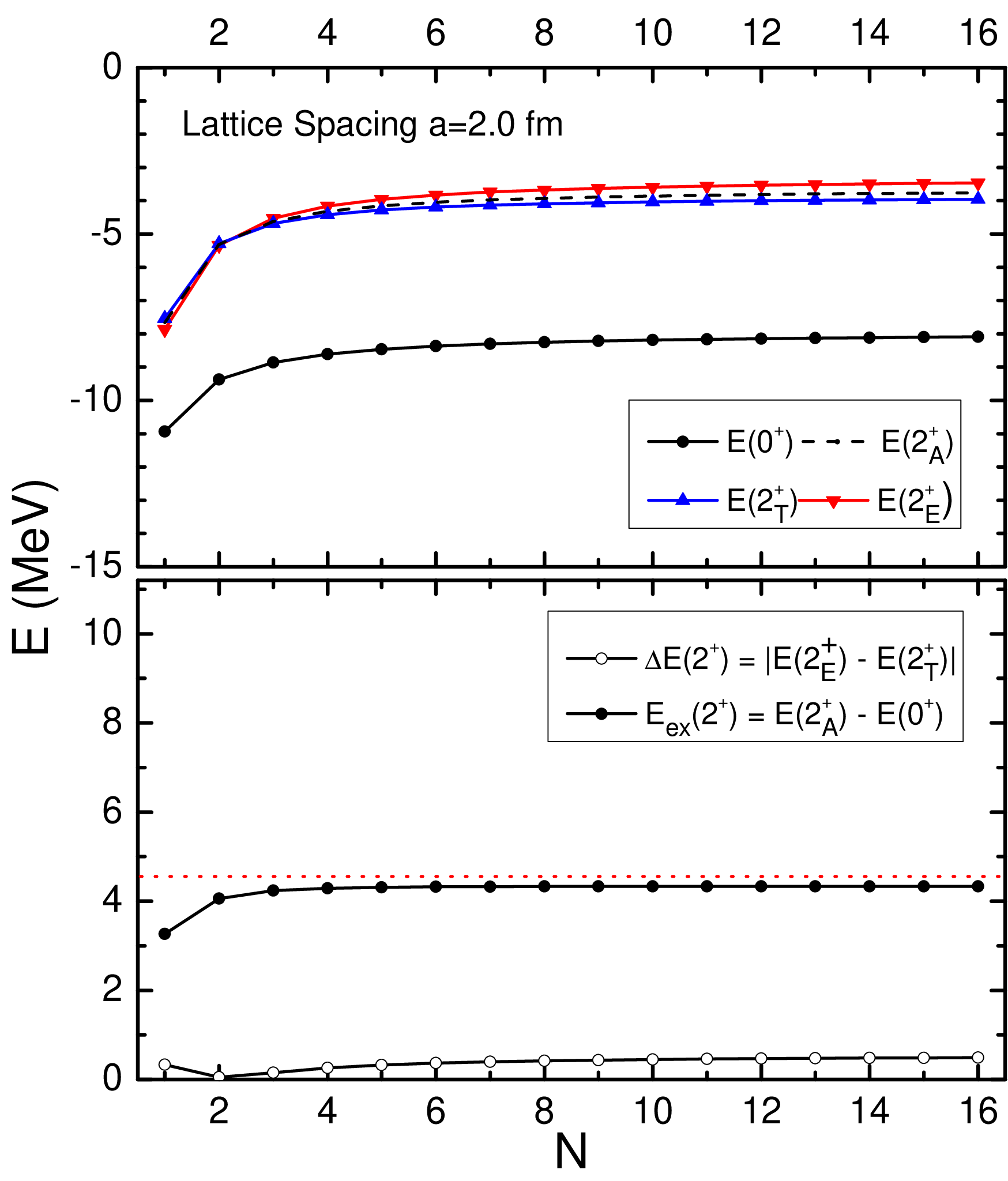}
  \includegraphics[width=0.49\columnwidth]{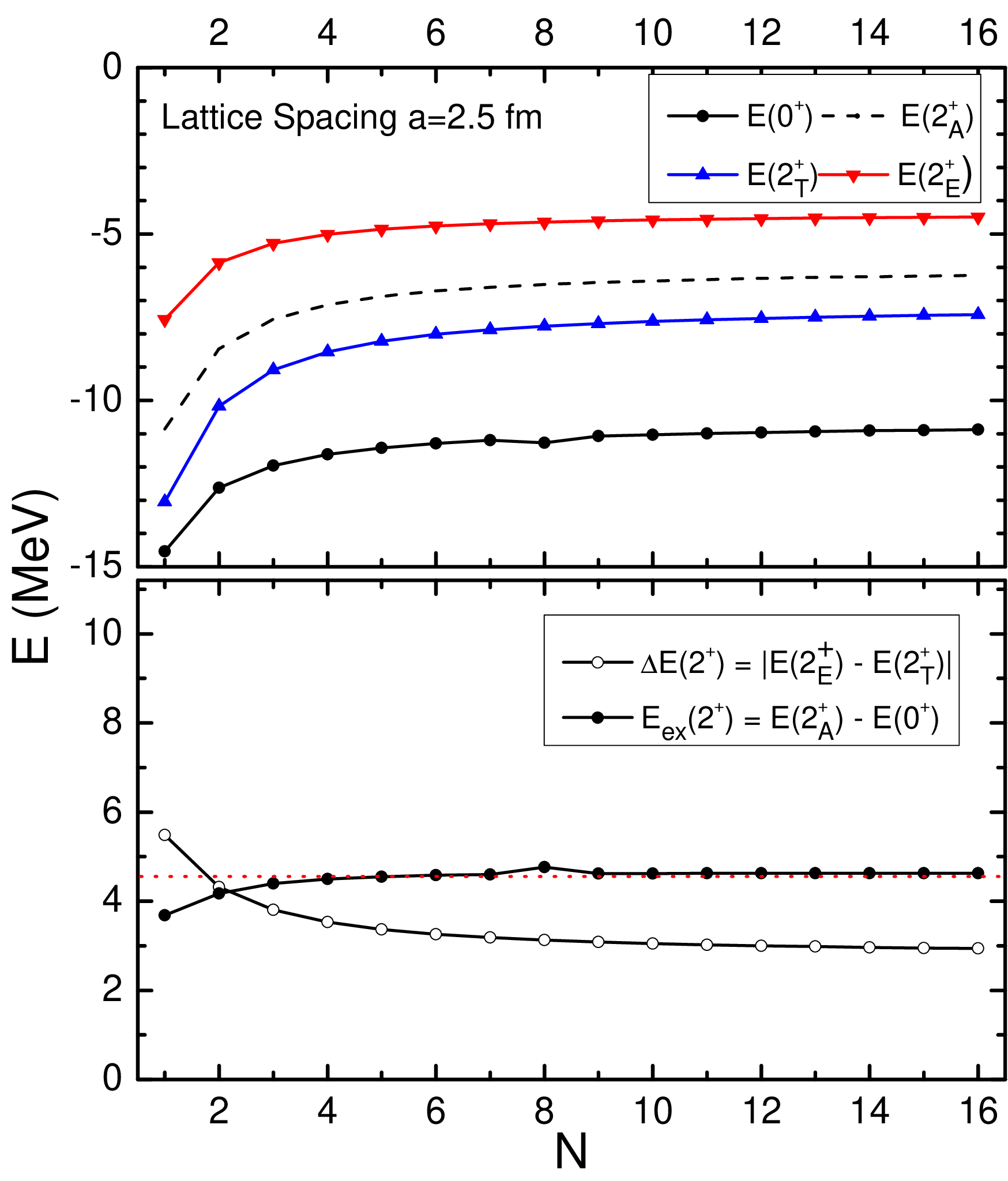}
  \caption{{\small \label{fig:The-same-as-2}\textit{Upper panels of each subfigure}{\small :
  Lowest $0^{+}$ and $2^{+}$ levels of $^{12}$C nucleus versus lattice improvement index $N$ for lattice spacings 
    $a = 1.2, 1.6, 2.0, 2.5~{\rm fm}$.  
  The box size $L$ is kept larger than 18 fm to remove finite volume effects.  
  The dashed lines represent the weighted averaged value $E(2_{{\rm A}}^{+})$.
  }\textit{\small Lower panels of each subfigure}{\small : 
  The energy splitting $\Delta E(2^+)$ (open circles) and the average excitation gap $E_{{\rm ex}}(2^{+})$ (full circles)  
    versus lattice improvement index $N$ for lattice spacings $a = 1.2, 1.6, 2.0, 2.5~{\rm fm}$.
	The dotted line shows the continuum infinite-volume limit result for $E_{{\rm ex}}(2^{+})$.
  }}}
\end{figure}

In Fig.~\ref{fig:The-same-as-2}, we show the $0^{+}$ and $2^{+}$ levels of $^{12}$C, $E(2_{{\rm A}}^{+})$, $E_{{\rm ex}}(2^{+})$,
  and $\Delta E(2^+)$ as functions of the derivative improvement index $N$ for lattice spacings $a = 1.2, 1.6, 2.0, 2.5~{\rm fm}$. 
In all cases the box size $L$ is taken to be $L\geq18$~fm.
In the lower panel of each subfigure, the dotted line shows the continuum infinite-volume limit result for $E_{{\rm ex}}(2^{+})$.
The results are qualitatively similar to those for the $^{8}$Be nucleus.
Although the errors in $E_{{\rm ex}}(2^{+})$ and $\Delta E(2^+)$ are not eliminated by the use of higher-order lattice actions
  for the kinetic energy, the results are clearly more accurate when using the improved actions.
As we found in the $^8$Be system, the fact that we are not including improvements to the interaction seem not so important 
  for the average excitation gap $E_{{\rm ex}}(2^{+})$. 
The error of the average excitation energy $E_{{\rm ex}}(2^{+})$ for $N\ge3$  comes very close to the continuum value,
  with an error of less than $0.5$~MeV for all lattice spacings considered.  
This process of calculating multiplet-averaged excitation gaps with improved dispersion actions can be used to decrease 
  lattice discretization errors even at large lattice spacing.

Our results are examined for $J=2$ orbits in both two-$\alpha$ and three-$\alpha$ systems. 
It is straightforward to extend them to high-$J$ levels and clustering systems with more $\alpha$ particles.
Furthermore, they can be applied to \textit{ab inito} calculations based on effective field theories with lattice simulations.
For example, the various branches from high-$J$ orbits may obey the rule discussed in this paper and the weighted average energy
  levels can serve as rather good approximations for the converged ones.

\section{Summary and discussion}

In any lattice calculation, finite volume effects and lattice discretization errors can induce unphysical shifts on energy levels.
Furthermore, the degeneracy of multiplets with the same angular momentum can be broken according to various representations of the
  cubic group. 
In this paper, we have employed an $\alpha$ cluster model in order to investigate the dependence of the unphysical splittings on 
  the volume and the lattice spacing.
We have shown that for both the $^{8}$Be and $^{12}$C nuclei, the splitting of the $2^+$ states depends on the lattice spacing
  and the intrinsic length scales in the system.
The energy for a given cubic represenation is minimized when the natural separation between particles is commensurate with the
  separation between lattice points along the preferred lattice directions. 

We have also shown that calculating multiplet-averaged excitation gaps with improved dispersion actions can be used to reduce 
  both finite volume effects and lattice discretization errors, even at large lattice spacing.
We have only considered $J=0$ and $J=2$ states, but we expect that our analysis should also apply to other values of $J$.  
This will be investigated in future work~\cite{Lu2014_inprogress}.

Although we considered a simple $\alpha$ cluster model in our analysis, the results should have immediate applications for 
  \textit{ab initio}  simulations of nuclei.
In systems where $\alpha$ clusters are believed to be an important part of the nuclear stucture, one can measure the $\alpha$ 
  dispersion relation.
Then, if necessary, one can try to correct for errors in the $\alpha$ dispersion relation via the underlying nucleon lattice 
  dispersion relation. The lesson we draw out of the analysis here, is that doing this may improve the spectrum of larger nuclei
  composed of $\alpha$ clusters. 
Also, the process of calculating multiplet averages and excitation gaps is straightforward, and this should certainly be 
considered when possible in future \textit{ab initio} lattice simulations.

We thank Thomas Luu for helpful discussions.
We acknowledge partial financial support from the 
Deutsche Forschungsgemeinschaft (Sino-German CRC 110), the Helmholtz Association (Contract No.\ VH-VI-417), 
BMBF (Grant No.\  05P12PDTFE),  the U.S. Department of Energy (DE-FG02-03ER41260), and
by the EU HadronPhysics3 project. 

\bibliographystyle{apsrev4-1}

\begin{thebibliography}{45}%
\makeatletter
\providecommand \@ifxundefined [1]{%
 \@ifx{#1\undefined}
}%
\providecommand \@ifnum [1]{%
 \ifnum #1\expandafter \@firstoftwo
 \else \expandafter \@secondoftwo
 \fi
}%
\providecommand \@ifx [1]{%
 \ifx #1\expandafter \@firstoftwo
 \else \expandafter \@secondoftwo
 \fi
}%
\providecommand \natexlab [1]{#1}%
\providecommand \enquote  [1]{``#1''}%
\providecommand \bibnamefont  [1]{#1}%
\providecommand \bibfnamefont [1]{#1}%
\providecommand \citenamefont [1]{#1}%
\providecommand \href@noop [0]{\@secondoftwo}%
\providecommand \href [0]{\begingroup \@sanitize@url \@href}%
\providecommand \@href[1]{\@@startlink{#1}\@@href}%
\providecommand \@@href[1]{\endgroup#1\@@endlink}%
\providecommand \@sanitize@url [0]{\catcode `\\12\catcode `\$12\catcode
  `\&12\catcode `\#12\catcode `\^12\catcode `\_12\catcode `\%12\relax}%
\providecommand \@@startlink[1]{}%
\providecommand \@@endlink[0]{}%
\providecommand \url  [0]{\begingroup\@sanitize@url \@url }%
\providecommand \@url [1]{\endgroup\@href {#1}{\urlprefix }}%
\providecommand \urlprefix  [0]{URL }%
\providecommand \Eprint [0]{\href }%
\providecommand \doibase [0]{http://dx.doi.org/}%
\providecommand \selectlanguage [0]{\@gobble}%
\providecommand \bibinfo  [0]{\@secondoftwo}%
\providecommand \bibfield  [0]{\@secondoftwo}%
\providecommand \translation [1]{[#1]}%
\providecommand \BibitemOpen [0]{}%
\providecommand \bibitemStop [0]{}%
\providecommand \bibitemNoStop [0]{.\EOS\space}%
\providecommand \EOS [0]{\spacefactor3000\relax}%
\providecommand \BibitemShut  [1]{\csname bibitem#1\endcsname}%
\let\auto@bib@innerbib\@empty
\bibitem [{\citenamefont {Lee}(2009)}]{Lee2009_PPNP63-117}%
  \BibitemOpen
  \bibfield  {author} {\bibinfo {author} {\bibfnamefont {D.}~\bibnamefont
  {Lee}},\ }\href {\doibase 10.1016/j.ppnp.2008.12.001} {\bibfield  {journal}
  {\bibinfo  {journal} {Prog. Part. Nucl. Phys.}\ }\textbf {\bibinfo {volume}
  {63}},\ \bibinfo {pages} {117} (\bibinfo {year} {2009})}\BibitemShut
  {NoStop}%
\bibitem [{\citenamefont {Bazavov}\ \emph {et~al.}(2010)\citenamefont
  {Bazavov}, \citenamefont {Toussaint}, \citenamefont {Bernard}, \citenamefont
  {Laiho}, \citenamefont {DeTar}, \citenamefont {Levkova}, \citenamefont
  {Oktay}, \citenamefont {Gottlieb}, \citenamefont {Heller}, \citenamefont
  {Hetrick}, \citenamefont {Mackenzie}, \citenamefont {Sugar},\ and\
  \citenamefont {Van~de Water}}]{Bazavov2010_RMP82-1349}%
  \BibitemOpen
  \bibfield  {author} {\bibinfo {author} {\bibfnamefont {A.}~\bibnamefont
  {Bazavov}}, \bibinfo {author} {\bibfnamefont {D.}~\bibnamefont {Toussaint}},
  \bibinfo {author} {\bibfnamefont {C.}~\bibnamefont {Bernard}}, \bibinfo
  {author} {\bibfnamefont {J.}~\bibnamefont {Laiho}}, \bibinfo {author}
  {\bibfnamefont {C.}~\bibnamefont {DeTar}}, \bibinfo {author} {\bibfnamefont
  {L.}~\bibnamefont {Levkova}}, \bibinfo {author} {\bibfnamefont {M.~B.}\
  \bibnamefont {Oktay}}, \bibinfo {author} {\bibfnamefont {S.}~\bibnamefont
  {Gottlieb}}, \bibinfo {author} {\bibfnamefont {U.~M.}\ \bibnamefont
  {Heller}}, \bibinfo {author} {\bibfnamefont {J.~E.}\ \bibnamefont {Hetrick}},
  \bibinfo {author} {\bibfnamefont {P.~B.}\ \bibnamefont {Mackenzie}}, \bibinfo
  {author} {\bibfnamefont {R.}~\bibnamefont {Sugar}}, \ and\ \bibinfo {author}
  {\bibfnamefont {R.~S.}\ \bibnamefont {Van~de Water}},\ }\href {\doibase
  10.1103/RevModPhys.82.1349} {\bibfield  {journal} {\bibinfo  {journal} {Rev.
  Mod. Phys.}\ }\textbf {\bibinfo {volume} {82}},\ \bibinfo {pages} {1349}
  (\bibinfo {year} {2010})}\BibitemShut {NoStop}%
\bibitem [{\citenamefont {Beane}\ \emph {et~al.}(2011)\citenamefont {Beane},
  \citenamefont {Detmold}, \citenamefont {Orginos},\ and\ \citenamefont
  {Savage}}]{Beane2011_PPNP66-1}%
  \BibitemOpen
  \bibfield  {author} {\bibinfo {author} {\bibfnamefont {S.}~\bibnamefont
  {Beane}}, \bibinfo {author} {\bibfnamefont {W.}~\bibnamefont {Detmold}},
  \bibinfo {author} {\bibfnamefont {K.}~\bibnamefont {Orginos}}, \ and\
  \bibinfo {author} {\bibfnamefont {M.}~\bibnamefont {Savage}},\ }\href
  {\doibase 10.1016/j.ppnp.2010.08.002} {\bibfield  {journal} {\bibinfo
  {journal} {Prog. Part. Nucl. Phys.}\ }\textbf {\bibinfo {volume} {66}},\
  \bibinfo {pages} {1} (\bibinfo {year} {2011})}\BibitemShut {NoStop}%
\bibitem [{\citenamefont {Borasoy}\ \emph {et~al.}(2007)\citenamefont
  {Borasoy}, \citenamefont {Epelbaum}, \citenamefont {Krebs}, \citenamefont
  {Lee},\ and\ \citenamefont {Mei{\ss}ner}}]{Borasoy2007_EPJA31-105}%
  \BibitemOpen
  \bibfield  {author} {\bibinfo {author} {\bibfnamefont {B.}~\bibnamefont
  {Borasoy}}, \bibinfo {author} {\bibfnamefont {E.}~\bibnamefont {Epelbaum}},
  \bibinfo {author} {\bibfnamefont {H.}~\bibnamefont {Krebs}}, \bibinfo
  {author} {\bibfnamefont {D.}~\bibnamefont {Lee}}, \ and\ \bibinfo {author}
  {\bibfnamefont {U.-G.}\ \bibnamefont {Mei{\ss}ner}},\ }\href {\doibase
  10.1140/epja/i2006-10154-1} {\bibfield  {journal} {\bibinfo  {journal} {Eur.
  Phys. J. A}\ }\textbf {\bibinfo {volume} {31}},\ \bibinfo {pages} {105}
  (\bibinfo {year} {2007})}\BibitemShut {NoStop}%
\bibitem [{\citenamefont {Epelbaum}\ \emph
  {et~al.}(2010{\natexlab{a}})\citenamefont {Epelbaum}, \citenamefont {Krebs},
  \citenamefont {Lee},\ and\ \citenamefont
  {Mei{\ss}ner}}]{Epelbaum2010_EPJA45-335}%
  \BibitemOpen
  \bibfield  {author} {\bibinfo {author} {\bibfnamefont {E.}~\bibnamefont
  {Epelbaum}}, \bibinfo {author} {\bibfnamefont {H.}~\bibnamefont {Krebs}},
  \bibinfo {author} {\bibfnamefont {D.}~\bibnamefont {Lee}}, \ and\ \bibinfo
  {author} {\bibfnamefont {U.-G.}\ \bibnamefont {Mei{\ss}ner}},\ }\href
  {\doibase 10.1140/epja/i2010-11009-x} {\bibfield  {journal} {\bibinfo
  {journal} {Eur. Phys. J. A}\ }\textbf {\bibinfo {volume} {45}},\ \bibinfo
  {pages} {335} (\bibinfo {year} {2010}{\natexlab{a}})}\BibitemShut {NoStop}%
\bibitem [{\citenamefont {Epelbaum}\ \emph
  {et~al.}(2010{\natexlab{b}})\citenamefont {Epelbaum}, \citenamefont {Krebs},
  \citenamefont {Lee},\ and\ \citenamefont
  {Mei{\ss}ner}}]{Epelbaum2010_PRL104-142501}%
  \BibitemOpen
  \bibfield  {author} {\bibinfo {author} {\bibfnamefont {E.}~\bibnamefont
  {Epelbaum}}, \bibinfo {author} {\bibfnamefont {H.}~\bibnamefont {Krebs}},
  \bibinfo {author} {\bibfnamefont {D.}~\bibnamefont {Lee}}, \ and\ \bibinfo
  {author} {\bibfnamefont {U.-G.}\ \bibnamefont {Mei{\ss}ner}},\ }\href
  {\doibase 10.1103/PhysRevLett.104.142501} {\bibfield  {journal} {\bibinfo
  {journal} {Phys. Rev. Lett.}\ }\textbf {\bibinfo {volume} {104}},\ \bibinfo
  {pages} {142501} (\bibinfo {year} {2010}{\natexlab{b}})}\BibitemShut
  {NoStop}%
\bibitem [{\citenamefont {Epelbaum}\ \emph {et~al.}(2011)\citenamefont
  {Epelbaum}, \citenamefont {Krebs}, \citenamefont {Lee},\ and\ \citenamefont
  {Mei{\ss}ner}}]{Epelbaum2011_PRL106-192501}%
  \BibitemOpen
  \bibfield  {author} {\bibinfo {author} {\bibfnamefont {E.}~\bibnamefont
  {Epelbaum}}, \bibinfo {author} {\bibfnamefont {H.}~\bibnamefont {Krebs}},
  \bibinfo {author} {\bibfnamefont {D.}~\bibnamefont {Lee}}, \ and\ \bibinfo
  {author} {\bibfnamefont {U.-G.}\ \bibnamefont {Mei{\ss}ner}},\ }\href
  {\doibase 10.1103/PhysRevLett.106.192501} {\bibfield  {journal} {\bibinfo
  {journal} {Phys. Rev. Lett.}\ }\textbf {\bibinfo {volume} {106}},\ \bibinfo
  {pages} {192501} (\bibinfo {year} {2011})}\BibitemShut {NoStop}%
\bibitem [{\citenamefont {Epelbaum}\ \emph {et~al.}(2013)\citenamefont
  {Epelbaum}, \citenamefont {Krebs}, \citenamefont {L{\"a}hde}, \citenamefont
  {Lee},\ and\ \citenamefont {Mei{\ss}ner}}]{Epelbaum2013_PRL110-112502}%
  \BibitemOpen
  \bibfield  {author} {\bibinfo {author} {\bibfnamefont {E.}~\bibnamefont
  {Epelbaum}}, \bibinfo {author} {\bibfnamefont {H.}~\bibnamefont {Krebs}},
  \bibinfo {author} {\bibfnamefont {T.~A.}\ \bibnamefont {L{\"a}hde}}, \bibinfo
  {author} {\bibfnamefont {D.}~\bibnamefont {Lee}}, \ and\ \bibinfo {author}
  {\bibfnamefont {U.-G.}\ \bibnamefont {Mei{\ss}ner}},\ }\href {\doibase
  10.1103/PhysRevLett.110.112502} {\bibfield  {journal} {\bibinfo  {journal}
  {Phys. Rev. Lett.}\ }\textbf {\bibinfo {volume} {110}},\ \bibinfo {pages}
  {112502} (\bibinfo {year} {2013})}\BibitemShut {NoStop}%
\bibitem [{\citenamefont {Johnson}(1982)}]{Johnson1982_PLB114-147}%
  \BibitemOpen
  \bibfield  {author} {\bibinfo {author} {\bibfnamefont {R.}~\bibnamefont
  {Johnson}},\ }\href {\doibase 10.1016/0370-2693(82)90134-4} {\bibfield
  {journal} {\bibinfo  {journal} {Phys. Lett. B}\ }\textbf {\bibinfo {volume}
  {114}},\ \bibinfo {pages} {147} (\bibinfo {year} {1982})}\BibitemShut
  {NoStop}%
\bibitem [{\citenamefont {Berg}\ and\ \citenamefont
  {Billoire}(1983)}]{Berg1983_NPB221-109}%
  \BibitemOpen
  \bibfield  {author} {\bibinfo {author} {\bibfnamefont {B.}~\bibnamefont
  {Berg}}\ and\ \bibinfo {author} {\bibfnamefont {A.}~\bibnamefont
  {Billoire}},\ }\href {\doibase 10.1016/0550-3213(83)90620-X} {\bibfield
  {journal} {\bibinfo  {journal} {Nucl. Phys. B}\ }\textbf {\bibinfo {volume}
  {221}},\ \bibinfo {pages} {109 } (\bibinfo {year} {1983})}\BibitemShut
  {NoStop}%
\bibitem [{\citenamefont {Mandula}\ \emph {et~al.}(1983)\citenamefont
  {Mandula}, \citenamefont {Zweig},\ and\ \citenamefont
  {Govaerts}}]{Mandula1983_NPB228-91}%
  \BibitemOpen
  \bibfield  {author} {\bibinfo {author} {\bibfnamefont {J.~E.}\ \bibnamefont
  {Mandula}}, \bibinfo {author} {\bibfnamefont {G.}~\bibnamefont {Zweig}}, \
  and\ \bibinfo {author} {\bibfnamefont {J.}~\bibnamefont {Govaerts}},\ }\href
  {\doibase 10.1016/0550-3213(83)90399-1} {\bibfield  {journal} {\bibinfo
  {journal} {Nucl. Phys. B}\ }\textbf {\bibinfo {volume} {228}},\ \bibinfo
  {pages} {91 } (\bibinfo {year} {1983})}\BibitemShut {NoStop}%
\bibitem [{\citenamefont {Baye}\ and\ \citenamefont
  {Heenen}(1984)}]{Baye1984_PRC29-1056}%
  \BibitemOpen
  \bibfield  {author} {\bibinfo {author} {\bibfnamefont {D.}~\bibnamefont
  {Baye}}\ and\ \bibinfo {author} {\bibfnamefont {P.-H.}\ \bibnamefont
  {Heenen}},\ }\href {\doibase 10.1103/PhysRevC.29.1056} {\bibfield  {journal}
  {\bibinfo  {journal} {Phys. Rev. C}\ }\textbf {\bibinfo {volume} {29}},\
  \bibinfo {pages} {1056} (\bibinfo {year} {1984})}\BibitemShut {NoStop}%
\bibitem [{\citenamefont {Dudek}\ \emph {et~al.}(2009)\citenamefont {Dudek},
  \citenamefont {Edwards}, \citenamefont {Peardon}, \citenamefont {Richards},\
  and\ \citenamefont {Thomas}}]{Dudek2009_PRL103-262001}%
  \BibitemOpen
  \bibfield  {author} {\bibinfo {author} {\bibfnamefont {J.~J.}\ \bibnamefont
  {Dudek}}, \bibinfo {author} {\bibfnamefont {R.~G.}\ \bibnamefont {Edwards}},
  \bibinfo {author} {\bibfnamefont {M.~J.}\ \bibnamefont {Peardon}}, \bibinfo
  {author} {\bibfnamefont {D.~G.}\ \bibnamefont {Richards}}, \ and\ \bibinfo
  {author} {\bibfnamefont {C.~E.}\ \bibnamefont {Thomas}} (\bibinfo
  {collaboration} {for the Hadron Spectrum Collaboration}),\ }\href {\doibase
  10.1103/PhysRevLett.103.262001} {\bibfield  {journal} {\bibinfo  {journal}
  {Phys. Rev. Lett.}\ }\textbf {\bibinfo {volume} {103}},\ \bibinfo {pages}
  {262001} (\bibinfo {year} {2009})}\BibitemShut {NoStop}%
\bibitem [{\citenamefont {Dudek}\ \emph {et~al.}(2010)\citenamefont {Dudek},
  \citenamefont {Edwards}, \citenamefont {Peardon}, \citenamefont {Richards},\
  and\ \citenamefont {Thomas}}]{Dudek2010_PRD82-034508}%
  \BibitemOpen
  \bibfield  {author} {\bibinfo {author} {\bibfnamefont {J.~J.}\ \bibnamefont
  {Dudek}}, \bibinfo {author} {\bibfnamefont {R.~G.}\ \bibnamefont {Edwards}},
  \bibinfo {author} {\bibfnamefont {M.~J.}\ \bibnamefont {Peardon}}, \bibinfo
  {author} {\bibfnamefont {D.~G.}\ \bibnamefont {Richards}}, \ and\ \bibinfo
  {author} {\bibfnamefont {C.~E.}\ \bibnamefont {Thomas}} (\bibinfo
  {collaboration} {for the Hadron Spectrum Collaboration}),\ }\href {\doibase
  10.1103/PhysRevD.82.034508} {\bibfield  {journal} {\bibinfo  {journal} {Phys.
  Rev. D}\ }\textbf {\bibinfo {volume} {82}},\ \bibinfo {pages} {034508}
  (\bibinfo {year} {2010})}\BibitemShut {NoStop}%
\bibitem [{\citenamefont {Edwards}\ \emph {et~al.}(2011)\citenamefont
  {Edwards}, \citenamefont {Dudek}, \citenamefont {Richards},\ and\
  \citenamefont {Wallace}}]{Edwards2011_PRD84-074508}%
  \BibitemOpen
  \bibfield  {author} {\bibinfo {author} {\bibfnamefont {R.~G.}\ \bibnamefont
  {Edwards}}, \bibinfo {author} {\bibfnamefont {J.~J.}\ \bibnamefont {Dudek}},
  \bibinfo {author} {\bibfnamefont {D.~G.}\ \bibnamefont {Richards}}, \ and\
  \bibinfo {author} {\bibfnamefont {S.~J.}\ \bibnamefont {Wallace}},\ }\href
  {\doibase 10.1103/PhysRevD.84.074508} {\bibfield  {journal} {\bibinfo
  {journal} {Phys. Rev. D}\ }\textbf {\bibinfo {volume} {84}},\ \bibinfo
  {pages} {074508} (\bibinfo {year} {2011})}\BibitemShut {NoStop}%
\bibitem [{\citenamefont {Meinel}(2012)}]{Meinel2012_PRD85-114510}%
  \BibitemOpen
  \bibfield  {author} {\bibinfo {author} {\bibfnamefont {S.}~\bibnamefont
  {Meinel}},\ }\href {\doibase 10.1103/PhysRevD.85.114510} {\bibfield
  {journal} {\bibinfo  {journal} {Phys. Rev. D}\ }\textbf {\bibinfo {volume}
  {85}},\ \bibinfo {pages} {114510} (\bibinfo {year} {2012})}\BibitemShut
  {NoStop}%
\bibitem [{\citenamefont {Davoudi}\ and\ \citenamefont
  {Savage}(2012)}]{Davoudi2012_PRD86-054505}%
  \BibitemOpen
  \bibfield  {author} {\bibinfo {author} {\bibfnamefont {Z.}~\bibnamefont
  {Davoudi}}\ and\ \bibinfo {author} {\bibfnamefont {M.~J.}\ \bibnamefont
  {Savage}},\ }\href {\doibase 10.1103/PhysRevD.86.054505} {\bibfield
  {journal} {\bibinfo  {journal} {Phys. Rev. D}\ }\textbf {\bibinfo {volume}
  {86}},\ \bibinfo {pages} {054505} (\bibinfo {year} {2012})}\BibitemShut
  {NoStop}%
\bibitem [{\citenamefont {L\"uscher}(1986)}]{Luescher1986_CMP104-177}%
  \BibitemOpen
  \bibfield  {author} {\bibinfo {author} {\bibfnamefont {M.}~\bibnamefont
  {L\"uscher}},\ }\href {\doibase 10.1007/BF01211589} {\bibfield  {journal}
  {\bibinfo  {journal} {Comm. Math. Phys.}\ }\textbf {\bibinfo {volume}
  {104}},\ \bibinfo {pages} {177} (\bibinfo {year} {1986})}\BibitemShut
  {NoStop}%
\bibitem [{\citenamefont {L\"uscher}(1991)}]{Luescher1991_NPB354-531}%
  \BibitemOpen
  \bibfield  {author} {\bibinfo {author} {\bibfnamefont {M.}~\bibnamefont
  {L\"uscher}},\ }\href {\doibase 10.1016/0550-3213(91)90366-6} {\bibfield
  {journal} {\bibinfo  {journal} {Nucl. Phys. B}\ }\textbf {\bibinfo {volume}
  {354}},\ \bibinfo {pages} {531 } (\bibinfo {year} {1991})}\BibitemShut
  {NoStop}%
\bibitem [{\citenamefont {Beane}\ \emph {et~al.}(2004)\citenamefont {Beane},
  \citenamefont {Bedaque}, \citenamefont {Parre{\~n}o},\ and\ \citenamefont
  {Savage}}]{Beane2004_PLB585-106}%
  \BibitemOpen
  \bibfield  {author} {\bibinfo {author} {\bibfnamefont {S.}~\bibnamefont
  {Beane}}, \bibinfo {author} {\bibfnamefont {P.}~\bibnamefont {Bedaque}},
  \bibinfo {author} {\bibfnamefont {A.}~\bibnamefont {Parre{\~n}o}}, \ and\
  \bibinfo {author} {\bibfnamefont {M.}~\bibnamefont {Savage}},\ }\href
  {\doibase 10.1016/j.physletb.2004.02.007} {\bibfield  {journal} {\bibinfo
  {journal} {Phys. Lett. B}\ }\textbf {\bibinfo {volume} {585}},\ \bibinfo
  {pages} {106 } (\bibinfo {year} {2004})}\BibitemShut {NoStop}%
\bibitem [{\citenamefont {K\"onig}\ \emph {et~al.}(2011)\citenamefont
  {K\"onig}, \citenamefont {Lee},\ and\ \citenamefont
  {Hammer}}]{Koenig2011_PRL107-112001}%
  \BibitemOpen
  \bibfield  {author} {\bibinfo {author} {\bibfnamefont {S.}~\bibnamefont
  {K\"onig}}, \bibinfo {author} {\bibfnamefont {D.}~\bibnamefont {Lee}}, \ and\
  \bibinfo {author} {\bibfnamefont {H.-W.}\ \bibnamefont {Hammer}},\ }\href
  {\doibase 10.1103/PhysRevLett.107.112001} {\bibfield  {journal} {\bibinfo
  {journal} {Phys. Rev. Lett.}\ }\textbf {\bibinfo {volume} {107}},\ \bibinfo
  {pages} {112001} (\bibinfo {year} {2011})}\BibitemShut {NoStop}%
\bibitem [{\citenamefont {K\"onig}\ \emph {et~al.}(2012)\citenamefont
  {K\"onig}, \citenamefont {Lee},\ and\ \citenamefont
  {Hammer}}]{Koenig2012_AP327-1450}%
  \BibitemOpen
  \bibfield  {author} {\bibinfo {author} {\bibfnamefont {S.}~\bibnamefont
  {K\"onig}}, \bibinfo {author} {\bibfnamefont {D.}~\bibnamefont {Lee}}, \ and\
  \bibinfo {author} {\bibfnamefont {H.-W.}\ \bibnamefont {Hammer}},\ }\href
  {\doibase 10.1016/j.aop.2011.12.015} {\bibfield  {journal} {\bibinfo
  {journal} {Annals Phys.}\ }\textbf {\bibinfo {volume} {327}},\ \bibinfo
  {pages} {1450} (\bibinfo {year} {2012})},\ \Eprint
  {http://arxiv.org/abs/1109.4577} {arXiv:1109.4577 [hep-lat]} \BibitemShut
  {NoStop}%
\bibitem [{\citenamefont {Bour}\ \emph {et~al.}(2011)\citenamefont {Bour},
  \citenamefont {K{\"o}nig}, \citenamefont {Lee}, \citenamefont {Hammer},\ and\
  \citenamefont {Mei{\ss}ner}}]{Bour2011_PRD84-091503}%
  \BibitemOpen
  \bibfield  {author} {\bibinfo {author} {\bibfnamefont {S.}~\bibnamefont
  {Bour}}, \bibinfo {author} {\bibfnamefont {S.}~\bibnamefont {K{\"o}nig}},
  \bibinfo {author} {\bibfnamefont {D.}~\bibnamefont {Lee}}, \bibinfo {author}
  {\bibfnamefont {H.-W.}\ \bibnamefont {Hammer}}, \ and\ \bibinfo {author}
  {\bibfnamefont {U.-G.}\ \bibnamefont {Mei{\ss}ner}},\ }\href {\doibase
  10.1103/PhysRevD.84.091503} {\bibfield  {journal} {\bibinfo  {journal} {Phys.
  Rev.}\ }\textbf {\bibinfo {volume} {84}},\ \bibinfo {pages} {091503}
  (\bibinfo {year} {2011})},\ \Eprint {http://arxiv.org/abs/1107.1272}
  {arXiv:1107.1272 [nucl-th]} \BibitemShut {NoStop}%
\bibitem [{\citenamefont {Brice{\~n}o}\ \emph {et~al.}()\citenamefont
  {Brice{\~n}o}, \citenamefont {Davoudi}, \citenamefont {Luu},\ and\
  \citenamefont {Savage}}]{Briceno2013_arXiv1311-7686}%
  \BibitemOpen
  \bibfield  {author} {\bibinfo {author} {\bibfnamefont {R.~A.}\ \bibnamefont
  {Brice{\~n}o}}, \bibinfo {author} {\bibfnamefont {Z.}~\bibnamefont
  {Davoudi}}, \bibinfo {author} {\bibfnamefont {T.~C.}\ \bibnamefont {Luu}}, \
  and\ \bibinfo {author} {\bibfnamefont {M.~J.}\ \bibnamefont {Savage}},\
  }\href {http://arxiv.org/abs/1311.7686} {\ }\Eprint
  {http://arxiv.org/abs/1311.7686} {arXiv:1311.7686 [hep-lat]} \BibitemShut
  {NoStop}%
\bibitem [{\citenamefont {Kreuzer}\ and\ \citenamefont
  {Hammer}(2010)}]{Kreuzer2010_EPJA43-229}%
  \BibitemOpen
  \bibfield  {author} {\bibinfo {author} {\bibfnamefont {S.}~\bibnamefont
  {Kreuzer}}\ and\ \bibinfo {author} {\bibfnamefont {H.-W.}\ \bibnamefont
  {Hammer}},\ }\href {\doibase 10.1140/epja/i2010-10910-6} {\bibfield
  {journal} {\bibinfo  {journal} {Eur. Phys. J. A}\ }\textbf {\bibinfo {volume}
  {43}},\ \bibinfo {pages} {229} (\bibinfo {year} {2010})}\BibitemShut
  {NoStop}%
\bibitem [{\citenamefont {Kreuzer}\ and\ \citenamefont
  {Hammer}(2011)}]{Kreuzer2011_PLB694-424}%
  \BibitemOpen
  \bibfield  {author} {\bibinfo {author} {\bibfnamefont {S.}~\bibnamefont
  {Kreuzer}}\ and\ \bibinfo {author} {\bibfnamefont {H.-W.}\ \bibnamefont
  {Hammer}},\ }\href {\doibase 10.1016/j.physletb.2010.10.003} {\bibfield
  {journal} {\bibinfo  {journal} {Phys. Lett. B}\ }\textbf {\bibinfo {volume}
  {694}},\ \bibinfo {pages} {424 } (\bibinfo {year} {2011})}\BibitemShut
  {NoStop}%
\bibitem [{\citenamefont {Weisz}(1983)}]{Weisz1983_NPB212-1}%
  \BibitemOpen
  \bibfield  {author} {\bibinfo {author} {\bibfnamefont {P.}~\bibnamefont
  {Weisz}},\ }\href {\doibase 10.1016/0550-3213(83)90595-3} {\bibfield
  {journal} {\bibinfo  {journal} {Nucl. Phys. B}\ }\textbf {\bibinfo {volume}
  {212}},\ \bibinfo {pages} {1 } (\bibinfo {year} {1983})}\BibitemShut
  {NoStop}%
\bibitem [{\citenamefont {Weisz}\ and\ \citenamefont
  {Wohlert}(1984)}]{Weisz1984_NPB236-397}%
  \BibitemOpen
  \bibfield  {author} {\bibinfo {author} {\bibfnamefont {P.}~\bibnamefont
  {Weisz}}\ and\ \bibinfo {author} {\bibfnamefont {R.}~\bibnamefont
  {Wohlert}},\ }\href {\doibase 10.1016/0550-3213(84)90543-1} {\bibfield
  {journal} {\bibinfo  {journal} {Nucl. Phys. B}\ }\textbf {\bibinfo {volume}
  {236}},\ \bibinfo {pages} {397 } (\bibinfo {year} {1984})}\BibitemShut
  {NoStop}%
\bibitem [{\citenamefont {L\"uscher}\ and\ \citenamefont
  {Weisz}(1984)}]{Luescher1984_NPB240-349}%
  \BibitemOpen
  \bibfield  {author} {\bibinfo {author} {\bibfnamefont {M.}~\bibnamefont
  {L\"uscher}}\ and\ \bibinfo {author} {\bibfnamefont {P.}~\bibnamefont
  {Weisz}},\ }\href {\doibase 10.1016/0550-3213(84)90270-0} {\bibfield
  {journal} {\bibinfo  {journal} {Nucl. Phys. B}\ }\textbf {\bibinfo {volume}
  {240}},\ \bibinfo {pages} {349 } (\bibinfo {year} {1984})}\BibitemShut
  {NoStop}%
\bibitem [{\citenamefont {Curci}\ \emph {et~al.}(1983)\citenamefont {Curci},
  \citenamefont {Menotti},\ and\ \citenamefont
  {Paffuti}}]{Curci1983_PLB130-205}%
  \BibitemOpen
  \bibfield  {author} {\bibinfo {author} {\bibfnamefont {G.}~\bibnamefont
  {Curci}}, \bibinfo {author} {\bibfnamefont {P.}~\bibnamefont {Menotti}}, \
  and\ \bibinfo {author} {\bibfnamefont {G.}~\bibnamefont {Paffuti}},\ }\href
  {\doibase 10.1016/0370-2693(83)91043-2} {\bibfield  {journal} {\bibinfo
  {journal} {Phys. Lett. B}\ }\textbf {\bibinfo {volume} {130}},\ \bibinfo
  {pages} {205} (\bibinfo {year} {1983})}\BibitemShut {NoStop}%
\bibitem [{\citenamefont {Hamber}\ and\ \citenamefont
  {Wucur}(1983)}]{Hamber1983_PLB133-351}%
  \BibitemOpen
  \bibfield  {author} {\bibinfo {author} {\bibfnamefont {H.~W.}\ \bibnamefont
  {Hamber}}\ and\ \bibinfo {author} {\bibfnamefont {C.~M.}\ \bibnamefont
  {Wucur}},\ }\href {\doibase 10.1016/0370-2693(83)90162-4} {\bibfield
  {journal} {\bibinfo  {journal} {Phys. Lett. B}\ }\textbf {\bibinfo {volume}
  {133}},\ \bibinfo {pages} {351 } (\bibinfo {year} {1983})}\BibitemShut
  {NoStop}%
\bibitem [{\citenamefont {Eguchi}\ and\ \citenamefont
  {Kawamoto}(1984)}]{Eguchi1984_NPB237-609}%
  \BibitemOpen
  \bibfield  {author} {\bibinfo {author} {\bibfnamefont {T.}~\bibnamefont
  {Eguchi}}\ and\ \bibinfo {author} {\bibfnamefont {N.}~\bibnamefont
  {Kawamoto}},\ }\href {\doibase 10.1016/0550-3213(84)90010-5} {\bibfield
  {journal} {\bibinfo  {journal} {Nucl. Phys. B}\ }\textbf {\bibinfo {volume}
  {237}},\ \bibinfo {pages} {609 } (\bibinfo {year} {1984})}\BibitemShut
  {NoStop}%
\bibitem [{\citenamefont {Sheikholeslami}\ and\ \citenamefont
  {Wohlert}(1985)}]{Sheikholeslami1985_NPB259-572}%
  \BibitemOpen
  \bibfield  {author} {\bibinfo {author} {\bibfnamefont {B.}~\bibnamefont
  {Sheikholeslami}}\ and\ \bibinfo {author} {\bibfnamefont {R.}~\bibnamefont
  {Wohlert}},\ }\href {\doibase 10.1016/0550-3213(85)90002-1} {\bibfield
  {journal} {\bibinfo  {journal} {Nucl. Phys. B}\ }\textbf {\bibinfo {volume}
  {259}},\ \bibinfo {pages} {572 } (\bibinfo {year} {1985})}\BibitemShut
  {NoStop}%
\bibitem [{\citenamefont {von Oertzen}\ \emph {et~al.}(2006)\citenamefont {von
  Oertzen}, \citenamefont {Freer},\ and\ \citenamefont
  {Kanada-Enyo}}]{Oertzen2006_PR432-43}%
  \BibitemOpen
  \bibfield  {author} {\bibinfo {author} {\bibfnamefont {W.}~\bibnamefont {von
  Oertzen}}, \bibinfo {author} {\bibfnamefont {M.}~\bibnamefont {Freer}}, \
  and\ \bibinfo {author} {\bibfnamefont {Y.}~\bibnamefont {Kanada-Enyo}},\
  }\href {\doibase 10.1016/j.physrep.2006.07.001} {\bibfield  {journal}
  {\bibinfo  {journal} {Phys. Rep.}\ }\textbf {\bibinfo {volume} {432}},\
  \bibinfo {pages} {43 } (\bibinfo {year} {2006})}\BibitemShut {NoStop}%
\bibitem [{\citenamefont {Ebran}\ \emph {et~al.}(2012)\citenamefont {Ebran},
  \citenamefont {Khan}, \citenamefont {Nik\ifmmode \check{s}\else
  \v{s}\fi{}i\ifmmode~\acute{c}\else \'{c}\fi{}},\ and\ \citenamefont
  {Vretenar}}]{Ebran2012_Nature487-341}%
  \BibitemOpen
  \bibfield  {author} {\bibinfo {author} {\bibfnamefont {J.-P.}\ \bibnamefont
  {Ebran}}, \bibinfo {author} {\bibfnamefont {E.}~\bibnamefont {Khan}},
  \bibinfo {author} {\bibfnamefont {T.}~\bibnamefont {Nik\ifmmode
  \check{s}\else \v{s}\fi{}i\ifmmode~\acute{c}\else \'{c}\fi{}}}, \ and\
  \bibinfo {author} {\bibfnamefont {D.}~\bibnamefont {Vretenar}},\ }\href
  {\doibase 10.1038/nature11246} {\bibfield  {journal} {\bibinfo  {journal}
  {Nature}\ }\textbf {\bibinfo {volume} {487}},\ \bibinfo {pages} {341}
  (\bibinfo {year} {2012})}\BibitemShut {NoStop}%
\bibitem [{\citenamefont {Epelbaum}\ \emph {et~al.}(2012)\citenamefont
  {Epelbaum}, \citenamefont {Krebs}, \citenamefont {L{\"a}hde}, \citenamefont
  {Lee},\ and\ \citenamefont {Mei{\ss}ner}}]{Epelbaum2012_PRL109-252501}%
  \BibitemOpen
  \bibfield  {author} {\bibinfo {author} {\bibfnamefont {E.}~\bibnamefont
  {Epelbaum}}, \bibinfo {author} {\bibfnamefont {H.}~\bibnamefont {Krebs}},
  \bibinfo {author} {\bibfnamefont {T.~A.}\ \bibnamefont {L{\"a}hde}}, \bibinfo
  {author} {\bibfnamefont {D.}~\bibnamefont {Lee}}, \ and\ \bibinfo {author}
  {\bibfnamefont {U.-G.}\ \bibnamefont {Mei{\ss}ner}},\ }\href {\doibase
  10.1103/PhysRevLett.109.252501} {\bibfield  {journal} {\bibinfo  {journal}
  {Phys. Rev. Lett.}\ }\textbf {\bibinfo {volume} {109}},\ \bibinfo {pages}
  {252501} (\bibinfo {year} {2012})}\BibitemShut {NoStop}%
\bibitem [{\citenamefont {Ali}\ and\ \citenamefont
  {Bodmer}(1966)}]{Ali1966_NP80-99}%
  \BibitemOpen
  \bibfield  {author} {\bibinfo {author} {\bibfnamefont {S.}~\bibnamefont
  {Ali}}\ and\ \bibinfo {author} {\bibfnamefont {A.}~\bibnamefont {Bodmer}},\
  }\href {\doibase 10.1016/0029-5582(66)90829-7} {\bibfield  {journal}
  {\bibinfo  {journal} {Nucl. Phys.}\ }\textbf {\bibinfo {volume} {80}},\
  \bibinfo {pages} {99 } (\bibinfo {year} {1966})}\BibitemShut {NoStop}%
\bibitem [{\citenamefont {Portilho}\ and\ \citenamefont
  {Coon}(1979)}]{Portilho1979_ZPA290-93}%
  \BibitemOpen
  \bibfield  {author} {\bibinfo {author} {\bibfnamefont {O.}~\bibnamefont
  {Portilho}}\ and\ \bibinfo {author} {\bibfnamefont {S.}~\bibnamefont
  {Coon}},\ }\href {\doibase 10.1007/BF01408484} {\bibfield  {journal}
  {\bibinfo  {journal} {Z. Physik A}\ }\textbf {\bibinfo {volume} {290}},\
  \bibinfo {pages} {93} (\bibinfo {year} {1979})}\BibitemShut {NoStop}%
\bibitem [{\citenamefont {Schmid}(1980)}]{Schmid1980_ZPA297-105}%
  \BibitemOpen
  \bibfield  {author} {\bibinfo {author} {\bibfnamefont {E.~W.}\ \bibnamefont
  {Schmid}},\ }\href {\doibase 10.1007/BF01421466} {\bibfield  {journal}
  {\bibinfo  {journal} {Z. Physik A}\ }\textbf {\bibinfo {volume} {297}},\
  \bibinfo {pages} {105} (\bibinfo {year} {1980})}\BibitemShut {NoStop}%
\bibitem [{\citenamefont {Papp}\ and\ \citenamefont
  {Moszkowski}(2008)}]{Papp2008_MPLB22-2201}%
  \BibitemOpen
  \bibfield  {author} {\bibinfo {author} {\bibfnamefont {Z.}~\bibnamefont
  {Papp}}\ and\ \bibinfo {author} {\bibfnamefont {S.}~\bibnamefont
  {Moszkowski}},\ }\href {\doibase 10.1142/S0217984908016984} {\bibfield
  {journal} {\bibinfo  {journal} {Mod. Phys. Lett. B}\ }\textbf {\bibinfo
  {volume} {22}},\ \bibinfo {pages} {2201} (\bibinfo {year}
  {2008})}\BibitemShut {NoStop}%
\bibitem [{\citenamefont {Buck}\ \emph {et~al.}(1977)\citenamefont {Buck},
  \citenamefont {Friedrich},\ and\ \citenamefont
  {Wheatley}}]{Buck1977_NPA275-246}%
  \BibitemOpen
  \bibfield  {author} {\bibinfo {author} {\bibfnamefont {B.}~\bibnamefont
  {Buck}}, \bibinfo {author} {\bibfnamefont {H.}~\bibnamefont {Friedrich}}, \
  and\ \bibinfo {author} {\bibfnamefont {C.}~\bibnamefont {Wheatley}},\ }\href
  {\doibase 10.1016/0375-9474(77)90287-1} {\bibfield  {journal} {\bibinfo
  {journal} {Nucl. Phys. A}\ }\textbf {\bibinfo {volume} {275}},\ \bibinfo
  {pages} {246 } (\bibinfo {year} {1977})}\BibitemShut {NoStop}%
\bibitem [{\citenamefont {Tang}\ \emph {et~al.}(1978)\citenamefont {Tang},
  \citenamefont {Le{M}ere},\ and\ \citenamefont
  {Thompson}}]{Tang1978_PR47-167}%
  \BibitemOpen
  \bibfield  {author} {\bibinfo {author} {\bibfnamefont {Y.~C.}\ \bibnamefont
  {Tang}}, \bibinfo {author} {\bibfnamefont {M.}~\bibnamefont {Le{M}ere}}, \
  and\ \bibinfo {author} {\bibfnamefont {D.~R.}\ \bibnamefont {Thompson}},\
  }\href {\doibase 10.1016/0370-1573(78)90175-8} {\bibfield  {journal}
  {\bibinfo  {journal} {Phys. Rep.}\ }\textbf {\bibinfo {volume} {47}},\
  \bibinfo {pages} {167} (\bibinfo {year} {1978})}\BibitemShut {NoStop}%
\bibitem [{\citenamefont {Orabi}\ \emph {et~al.}(2008)\citenamefont {Orabi},
  \citenamefont {Suzuki}, \citenamefont {Matsumura}, \citenamefont {Fujiwara},
  \citenamefont {Baye}, \citenamefont {Descouvemont},\ and\ \citenamefont
  {Theeten}}]{Orabi2008_JPCS111-012045}%
  \BibitemOpen
  \bibfield  {author} {\bibinfo {author} {\bibfnamefont {M.}~\bibnamefont
  {Orabi}}, \bibinfo {author} {\bibfnamefont {Y.}~\bibnamefont {Suzuki}},
  \bibinfo {author} {\bibfnamefont {H.}~\bibnamefont {Matsumura}}, \bibinfo
  {author} {\bibfnamefont {Y.}~\bibnamefont {Fujiwara}}, \bibinfo {author}
  {\bibfnamefont {D.}~\bibnamefont {Baye}}, \bibinfo {author} {\bibfnamefont
  {P.}~\bibnamefont {Descouvemont}}, \ and\ \bibinfo {author} {\bibfnamefont
  {M.}~\bibnamefont {Theeten}},\ }\href {\doibase
  10.1088/1742-6596/111/1/012045} {\bibfield  {journal} {\bibinfo  {journal}
  {J. Phys.: Conference Series}\ }\textbf {\bibinfo {volume} {111}},\ \bibinfo
  {pages} {012045} (\bibinfo {year} {2008})}\BibitemShut {NoStop}%
\bibitem [{\citenamefont {Tursunov}(2001)}]{Tursunov2001_JPG27-1381}%
  \BibitemOpen
  \bibfield  {author} {\bibinfo {author} {\bibfnamefont {E.~M.}\ \bibnamefont
  {Tursunov}},\ }\href {\doibase 10.1088/0954-3899/27/7/301} {\bibfield
  {journal} {\bibinfo  {journal} {J. Phys. G}\ }\textbf {\bibinfo {volume}
  {27}},\ \bibinfo {pages} {1381} (\bibinfo {year} {2001})}\BibitemShut
  {NoStop}%
\bibitem [{\citenamefont {Lu}\ \emph {et~al.}()\citenamefont {Lu},
  \citenamefont {L{\"a}hde}, \citenamefont {Lee},\ and\ \citenamefont
  {Mei{\ss}ner}}]{Lu2014_inprogress}%
  \BibitemOpen
  \bibfield  {author} {\bibinfo {author} {\bibfnamefont {B.-N.}\ \bibnamefont
  {Lu}}, \bibinfo {author} {\bibfnamefont {T.~A.}\ \bibnamefont {L{\"a}hde}},
  \bibinfo {author} {\bibfnamefont {D.}~\bibnamefont {Lee}}, \ and\ \bibinfo
  {author} {\bibfnamefont {U.-G.}\ \bibnamefont {Mei{\ss}ner}},\ }\href@noop {}
  {\bibinfo  {journal} {{\it work in progress}}\ }\BibitemShut {NoStop}%
\end{thebibliography}
%

\end{document}